\documentclass[sigconf,table,usenames,dvipsnames,svgnames]{acmart}

\usepackage[np]{numprint}
\usepackage[utf8]{inputenc}
\usepackage[T1]{fontenc}
\usepackage{graphicx}
\usepackage{arydshln}
\usepackage{booktabs}
\usepackage{arydshln}
\usepackage{multirow}
\usepackage{import}
\usepackage{fp}
\usepackage{comment}
\usepackage{flushend}
\hypersetup{pdfstartview=FitV,
  breaklinks=true, colorlinks=true, linkcolor=black, citecolor=black, urlcolor=blue,
  pdftitle={Clusters in the Expanse: Understanding and Unbiasing IPv6 Hitlists}, pdfauthor={}, pdfkeywords={}
}
\usepackage{cleveref}
\usepackage{subcaption}
\usepackage{xspace}
\usepackage[binary-units]{siunitx}
\DeclareSIUnit{\packet}{p}
\usepackage[nolist]{acronym}
\usepackage{todonotes} %
\presetkeys{todonotes}{fancyline, color=olive!30}{}

\usepackage{pifont} %
\newcommand{\cmark}{\textcolor{PineGreen}{\ding{51}}}%
\newcommand{\tmark}{\textcolor{Orange}{\ding{109}}}%
\newcommand{\xmark}{\textcolor{Red}{\ding{55}}}%
\presetkeys%
{todonotes}%
{inline}{}
\usepackage[normalem]{ulem}
\usepackage{microtype}

\usepackage[absolute,showboxes]{textpos}

\newcommand{\zmapsix}{ZMapv6\xspace}
\newcommand{\eip}{Entropy/IP\xspace}
\newcommand{\sixgen}{6Gen\xspace}
\newcommand{\scamper}{scamper\xspace}

\newcommand{\eg}{e.g.,\xspace}
\newcommand{\ie}{i.e.,\xspace}
\newcommand{\etal}{et al.\xspace}
\newcommand{\zesplot}{zesplot\xspace}

\newcommand{\one}{(1)\xspace}
\newcommand{\two}{(2)\xspace}
\newcommand{\three}{(3)\xspace}
\newcommand{\four}{(4)\xspace}

\newcommand{\hitlist}{hitlist\xspace}
\newcommand{\hitlists}{hitlists\xspace}

\newif\ifcutoptional
\cutoptionalfalse
\DeclareSIUnit{\nothing}{\relax}
\newcommand{\gn}{\giga\nothing}
\newcommand{\mn}{\mega\nothing}
\newcommand{\kn}{\kilo\nothing}

\newcommand{\sk}[1]{\SI{#1}{\kn}}
\newcommand{\sm}[1]{\SI{#1}{\mn}}
\newcommand{\sg}[1]{\SI{#1}{\gn}}
\newcommand{\sperc}[1]{\SI{#1}{\percent}}
\newcommand{\sdollar}[1]{\SI{#1}[\$]{}}

\renewcommand*{\paragraph}[1]{%
    \vspace{.5em}
    \noindent
    {\normalfont \bf #1}
}

\usepackage{enumitem}
\setlist[itemize]{leftmargin=4mm}
\copyrightyear{2018}
\acmYear{2018}
\setcopyright{acmlicensed}
\acmConference[IMC '18]{2018 Internet Measurement Conference}{October 31-November 2, 2018}{Boston, MA, USA}
\acmBooktitle{2018 Internet Measurement Conference (IMC '18), October 31-November 2, 2018, Boston, MA, USA}
\acmPrice{15.00}
\acmDOI{10.1145/3278532.3278564}
\acmISBN{978-1-4503-5619-0/18/10}


\makeatother
\begin{document}

\title[Clusters in the Expanse: Understanding and Unbiasing IPv6 Hitlists]{Clusters in the Expanse: \\ Understanding and Unbiasing IPv6 Hitlists}

\author{Oliver Gasser}
    \affiliation{Technical University of Munich}
    \email{gasser@net.in.tum.de}
\author{Quirin Scheitle}
    \affiliation{Technical University of Munich}
    \email{scheitle@net.in.tum.de}
\author{Pawel Foremski}
    \affiliation{IITiS PAN} %
    \email{pjf@iitis.pl}
\author{Qasim Lone}
    \affiliation{Grenoble Alps University}
    \email{qasim.lone@univ-grenoble-alpes.fr}
\author{Maciej Korczy\'nski}
    \affiliation{Grenoble Alps University}
    \email{maciej.korczynski@univ-grenoble-alpes.fr}
\author{Stephen D. Strowes}
    \affiliation{RIPE NCC}
    \email{sdstrowes@gmail.com}
\author{Luuk Hendriks}
    \affiliation{University of Twente}
    \email{luuk.hendriks@utwente.nl}
\author{Georg Carle}
    \affiliation{Technical University of Munich}
    \email{carle@net.in.tum.de}


\begin{abstract}
Network measurements are an important tool in understanding the Internet.
Due to the expanse of the IPv6 address space, exhaustive scans as in IPv4 are not possible for IPv6. %
In recent years, several studies have proposed the use of target lists of IPv6 addresses, called IPv6 \hitlists.

In this paper, we show that addresses in IPv6 \hitlists are heavily clustered.
We present novel techniques that allow IPv6 \hitlists to be pushed from quantity to quality. %
We perform a longitudinal active measurement study over 6 months, targeting more than \SI{50}{\mn} addresses.
We develop a rigorous method to detect aliased prefixes, %
which identifies \sperc{1.5} of our prefixes as aliased, pertaining to about half of our target addresses.
Using entropy clustering, we group the entire hitlist into just 6 distinct addressing schemes. 
Furthermore, we perform client measurements by leveraging crowdsourcing.

To encourage reproducibility in network measurement research and to serve as a starting point for future IPv6 studies, we publish source code, analysis tools, and data. %
\end{abstract}

\begingroup
\mathchardef\UrlBreakPenalty=10000

\maketitle
\renewcommand{\shortauthors}{Gasser, Scheitle, Foremski, Lone, Korczynski, Strowes, Hendriks, and Carle}

\setlength{\TPHorizModule}{\paperwidth}
\setlength{\TPVertModule}{\paperheight}
\TPMargin{5pt}
\begin{textblock}{0.8}(0.1,0.02)
    \noindent
    \footnotesize
    If you cite this paper, please use the IMC reference:
    Oliver Gasser, Quirin Scheitle, Pawel Foremski, Qasim Lone, Maciej Korczyński, Stephen D. Strowes, Luuk Hendriks, and Georg Carle. 2018.
    Clusters in the Expanse:, Understanding and Unbiasing IPv6 Hitlists.
    In \textit{2018 Internet Measurement Conference (IMC '18), October 31-November 2, 2018, Boston, MA, USA.}
    ACM, New York, NY, USA, 15 pages.
    \url{https://doi.org/10.1145/3278532.3278564}
\end{textblock}

\section{Introduction}

Internet scanning has a rich history of generating insights for security, topology, routing, and many other fields. 
Advances in software and link speeds in recent years allow the entire IPv4 Internet to be easily scanned in just a few minutes~\cite{Durumeric2013, adrian2014zippier, massscan}.
However, scanning the expanse of the entire IPv6 Internet is infeasible due to its size, which is magnitudes above both what can technically be sent or stored, and what is an ethical volume of queries to be sent to a system or network. 
Therefore, state-of-the-art IPv6 Internet scanning resorts to the methods used in the early days of IPv4 Internet scanning, \ie using lists of target IP addresses, so-called \hitlists, which served as a representative subset of the IPv4 address space~\cite{fan2010selecting, claffy2009internet, bush2007testing}.

The IPv6 address space also comes with unique, different challenges to such \hitlists.
First, \hitlists can be biased (\ie not representative of the Internet as a whole) due to imbalanced Autonomous System (AS) and prefix representations or IP address aliasing.
Second, due to similarly large allocation sizes, a single network---or even a single machine\footnote{We have indeed seen a web server responding to an entire /32 prefix, \ie $2^{96}$ addresses.}---can easily overwhelm a hitlist with countless IP addresses. 
Third, addresses might be used only for very brief periods of time, as there is no pressure for re-use. 
Thus, a key quality of IPv6 \hitlists is not the count of IP addresses, but responsiveness and balance over ASes and prefixes. %
In this paper, we systematically tackle these challenges by:

\textbf{Comprehensive Address Discovery: }The first step in unbiasing a hitlist is creating a \emph{comprehensive} hitlist, for which we draw IP addresses from a multitude of state-of-the-art sources, cf. \Cref{sec:sources}.

\textbf{Clustering by Entropy: }To discover and understand clusters in the expanse of the IPv6 space, we leverage entropy analysis of IPv6 addresses. This helps to determine addressing schemes and aggregate clusters, which we explore in \Cref{sec:clustering:entropy}. 

\textbf{De-Aliasing: }To reduce the potential impact of aliased prefixes---\ie a single machine responding to all addresses in a possibly large prefix---we postulate and implement a rigorous method for aliased prefix detection, which we present in~\Cref{sec:alias}.

\textbf{Longitudinal Stability Probing: } To find reliably responsive addresses, we conduct longitudinal scans for our hitlist across several protocols. As expected, we find only a fraction of discovered IP addresses to actually respond to probing, which is an important filtering criterion for curation of an unbiased hitlist (cf. \Cref{sec:probing}).

\textbf{Diversifying Address Population: }We also evaluate three orthogonal methods to push the frontier of IPv6 hitlists: generating addresses using \eip~\cite{foremski2016entropy} and 6Gen~\cite{murdock2017target} (in~\Cref{sec:learning}), leveraging reverse DNS records~\cite{fiebig2017something} (in~\Cref{sec:rdns}), and crowdsourcing client IP addresses (in~\Cref{sec:client}).

\textbf{Plotting for Exploratory Analysis: }Visualizing IPv6 datasets is
challenging, as IPv4 approaches, such as describing the
entire address space with a Hilbert curve, do not scale for IPv6. We present a new plotting technique
that works with selective input, \eg prefixes announced in BGP, instead of
visualizing the entire address space.  We explain how to interpret these plots
in~\Cref{sec:sources}, and use it throughout the paper to give an intuitive
view of our data.
 
\textbf{Publication and Sharing: }Our open sharing of a reliable IPv6 hitlist has already supported various scientific studies~\cite{almeida2017characterization,Gasser2016b,scheitle2017hloc,gasser2017amplification,amann2017mission,siblings,beverly2018ip,foremski2016entropy,caastudy17,pam18ctlog,hendriks2017potential}.  %
We also publicly share daily snapshots of the curated and unbiased IPv6 hitlist created in this work. 
 
Throughout our work, we aim to adhere to highest of ethical standards (cf. \Cref{sec:ethical}), and aim for our work to be fully reproducible. 
We share code and data at:\\ %
\centerline{\url{https://ipv6hitlist.github.io}}

\section{Previous Work}
\begin{table}[b]
    \caption{Comparing this work with four previous works (ordered chronologically) based on the following metrics: number of addresses from public sources (\#publ.), BGP prefixes (\#pfx.), ASes; addresses from private sources (\#priv.); include client addresses (Cts), perform active probing (Prob.), perform aliased prefix detection (APD).} %
\label{tab:relwork}
\resizebox{\columnwidth}{!}{
    \begin{tabular}{l@{\hskip 1mm}r@{\hskip 2mm}r@{\hskip 2mm}r@{\hskip 2mm}r@{\hskip 2mm}c@{\hskip 2mm}c@{\hskip 2mm}c}%
	\toprule
    Previous work & \#publ. & \#pfx. & \#ASes & \#priv. & Cts & Prob. & APD \\ %
	\midrule
    Gasser \etal \cite{Gasser2016a}             & \sm{2.7} & \sk{5.8} & \sk{8.6} & \sm{149} & \cmark & \cmark & \xmark \\ %
    Foremski \etal \cite{foremski2016entropy}   & \sk{620} & <100\textsuperscript{1} & <100\textsuperscript{1} & \sg{3.5} & \cmark & \cmark & \xmark \\ %
    Fiebig \etal \cite{fiebig2017something} & \sm{2.8} & n/a\textsuperscript{2} & n/a & 0 & \cmark & \xmark & \xmark \\
    Murdock \etal \cite{murdock2017target}\textsuperscript{3} & \sm{1.0} & \sk{2.8} & \sk{2.4} & 0 & \cmark & \cmark & \tmark \\ %
	\midrule
    \textbf{This work}                          & \sm{55.1} & \sk{25.5} & \sk{10.9} & 0 & \cmark & \cmark & \cmark \\ %
	\bottomrule
    \multicolumn{8}{l}{\footnotesize 1: 15 networks, with few prefixes and ASes. 2: \sk{582} /64s. 3: Responsive addresses.}
\end{tabular}
}
\end{table}

Recent studies find that there are hundreds of million active IPv4 addresses \cite{scansio:icmpv4,Bano:2018:SIL:3213232.3213234,richter2016beyond,zander2014capturing}.
This densely populated IPv4 space is well suited for brute-force measurement approaches by scanning the complete address space \cite{Durumeric2013,adrian2014zippier}.
The IPv6 space, however, is extremely sparse. We survey previous work on targeting the sparse IPv6 space and compare our work with the state of the art in \Cref{tab:relwork}.

\textbf{DNS Techniques:}
The DNS was long known to be a possible source for IPv6 addresses \cite{vandijk2012,rfc7707}.
Strowes proposes using the \textit{in-addr.arpa} IPv4 rDNS tree to gather names that may be resolved to IPv6 addresses \cite{strowes2016}, on an assumption that naming is common between protocols.
This discovered \sk{965} IPv6 addresses located in some \num{5531} ASes, many of which (\sperc{56.7}) were responsive.
More recently, Fiebig \etal walked the rDNS tree to obtain \sm{2.8} IPv6 addresses \cite{fiebig2017something,fiebig2018rdns}.
They did not probe these addresses, leaving the question of responsiveness open.
Borgolte \etal find IPv6 addresses by NSEC-walking DNSSEC signed reverse zones~\cite{borgolte2018enumerating}.

In this work, we evaluate the responsiveness of IPv6 rDNS as a source and find rDNS IPv6 addresses a valuable addition to a hitlist.

\textbf{Structural Properties:}
The IPv6 address space may be sparsely populated, but address plans inside networks tend to indicate structure.
Recent related work leverages the structure of IPv6 addressing schemes to find new addresses.
Ullrich \etal use rule mining to find a few hundred IPv6 addresses \cite{ullrich2015reconnaissance}.
With \eip, Foremski \etal present a machine learning approach which trains on collected IPv6 addresses to build an addressing scheme model and generate new addresses \cite{foremski2016entropy}.
In this work, we refine \eip to generate IPv6 addresses for probing, and we introduce a new entropy clustering technique.
Similar to efforts leveraging dense address areas in IPv4 \cite{klwr-tbicr-16b}, Murdock \etal presented \sixgen to find dense regions in the IPv6 address space and generate neighboring addresses \cite{murdock2017target}.
In this work, we use \sixgen to generate addresses based on our hitlist and compare its performance with \eip.
Murdock \etal also performed a basic variant of aliased prefix detection (APD), which we extend in this work.
Plonka and Berger harness the structure from IPv6 address plans to allow large datasets to be shared~\cite{plonka:kip}.

\textbf{Hitlists:}
Gasser \etal~\cite{Gasser2016a} assemble a hitlist from a multitude of sources.
The vast majority of their \sm{149} IPv6 addresses, however, are obtained from non-public passive sources.
We build upon their approach, but exclusively use publicly available sources in order to make our work reproducible.
Recently, Beverly \etal analyze the IPv6 topology using large-scale traceroutes, leveraging Gasser \etal's public hitlist among other sources \cite{beverly2018ip}.

\textbf{Crowdsourcing:}
There are a few studies that have leveraged crowdsourcing platforms to perform network measurements~\cite{spoof,huz2015experience,varvello2016eyeorg}. 
We build on prior work by Huz \etal~\cite{huz2015experience}, who used the Amazon Mechanical Turk platform to test broadband speeds and IPv6 adoption. 
Compared to Huz \etal's 38 IPv6 addresses collected in 2015, we find many more.
\section{IPv6 Hitlist Sources}\label{sec:sources}

We leverage a variety of sources, for which we provide an overview in \Cref{tab:sources}.
Our guiding principle in selecting sources was that they should be \textit{public}, \ie accessible to anyone for free, in order to make our work reproducible and to allow fellow researchers to deploy variations of our IPv6 hitlist.
We consider data with an open and usually positive access decision process as public, such as Verisign's process to access zone files. 
We also aim to have balanced sources, which include servers, routers, and a share of clients. 
The sources we leverage are as follows:

\begin{table*}
	\centering
	\caption{Overview of hitlist sources, as of May 11, 2018.}
	\label{tab:sources}
    	\resizebox{\textwidth}{!}{
	\begin{tabular}{lrlrrrrrrr}
		\toprule
		Name & Public & Nature & IPs & new IPs & \#ASes & \#PFXes & Top AS1 & Top AS2 & Top AS3  \\
		\midrule
        DL: Domain Lists\textsuperscript{1}  & Yes & Servers & \sm{9.8} & \sm{9.8} & \sk{6.1} & \sk{10.3} & \sperc{89.7}\textcolor{BrickRed}{\ding{72}}&\sperc{2.0}\textcolor{BrickRed}{\ding{108}}&\sperc{1.5}\textcolor{BrickRed}{\ding{110}} \\ %
        FDNS: Rapid7 FDNS & Yes & Servers & \sm{3.3} & \sm{2.5} & \sk{7.7} & \sk{13.6} & \sperc{16.7}\textcolor{BrickRed}{\ding{72}}&\sperc{8.9}\textcolor{BrickRed}{\ding{115}}&\sperc{6.7}\textcolor{BrickRed}{\ding{67}} \\ %
        CT: Domains from CT logs\textsuperscript{2} & Yes & Servers & \sm{18.5} & \sm{16.2} & \sk{5.3} & \sk{8.7} & \sperc{92.3}\textcolor{BrickRed}{\ding{72}}&\sperc{1.6}\textcolor{BrickRed}{\ding{67}}&\sperc{0.8}\textcolor{blue}{\ding{72}} \\ %
        AXFR: AXFR\&TLDR & Yes & Mixed & \sm{0.7} & \sm{0.5} & \sk{3.2} & \sk{4.7} & \sperc{57.0}\textcolor{BrickRed}{\ding{72}}&\sperc{14.0}\textcolor{blue}{\ding{108}}&\sperc{8.3}\textcolor{BrickRed}{\ding{110}} \\ %
        BIT: Bitnodes & Yes & Mixed & \sk{31} & \sk{27} & 695 & \sk{1.4} & \sperc{8.0}\textcolor{blue}{\ding{72}}&\sperc{6.0}\textcolor{blue}{\ding{110}}&\sperc{6.0}\textcolor{blue}{\ding{115}} \\ %
        RA: RIPE Atlas\textsuperscript{3} & Yes & Routers & \sm{0.2} & \sm{0.2} & \sk{8.4} & \sk{19.1} & \sperc{6.6}\textcolor{BrickRed}{\ding{67}}&\sperc{3.5}\textcolor{BrickRed}{\ding{72}}&\sperc{3.1}\textcolor{blue}{\ding{67}} \\ %
        Scamper & -- & Routers &  \sm{26.0} & \sm{25.9} & \sk{6.3} & \sk{9.8} & \sperc{38.9}\textcolor{YellowOrange}{\ding{72}}&\sperc{23.8}\textcolor{YellowOrange}{\ding{108}}&\sperc{12.0}\textcolor{YellowOrange}{\ding{110}} \\ %
		\midrule 
        Total  & & & \sm{58.5} & \sm{55.1} & \sk{10.9} & \sk{25.5} & \sperc{45.4}\textcolor{BrickRed}{\ding{72}}&\sperc{18.4}\textcolor{YellowOrange}{\ding{72}}&\sperc{11.5}\textcolor{YellowOrange}{\ding{108}} \\
		\bottomrule
        \multicolumn{10}{l}{\footnotesize 1: Zone Files, Toplists, Blacklists (partially with NDA); 2: Excluding DNS names already included in Domain Lists; 3: Traceroute and ipmap data} \\
        \multicolumn{10}{l}{\footnotesize \textcolor{BrickRed}{\ding{72}}Amazon, \textcolor{BrickRed}{\ding{108}}Host Europe, \textcolor{BrickRed}{\ding{110}}Cloudflare, \textcolor{BrickRed}{\ding{115}}Linode, \textcolor{BrickRed}{\ding{67}}DTAG, \textcolor{blue}{\ding{72}}ProXad, \textcolor{blue}{\ding{108}}Hetzner, \textcolor{blue}{\ding{110}}Comcast, \textcolor{blue}{\ding{115}}Swisscom, \textcolor{blue}{\ding{67}}Google, \textcolor{YellowOrange}{\ding{72}}Antel, \textcolor{YellowOrange}{\ding{108}}Versatel, \textcolor{YellowOrange}{\ding{110}}BIHNET}
	\end{tabular}
}
\end{table*}
\begin{figure*}
	\centering
	\begin{subfigure}{0.3\textwidth}%
		\includegraphics[width=\columnwidth]{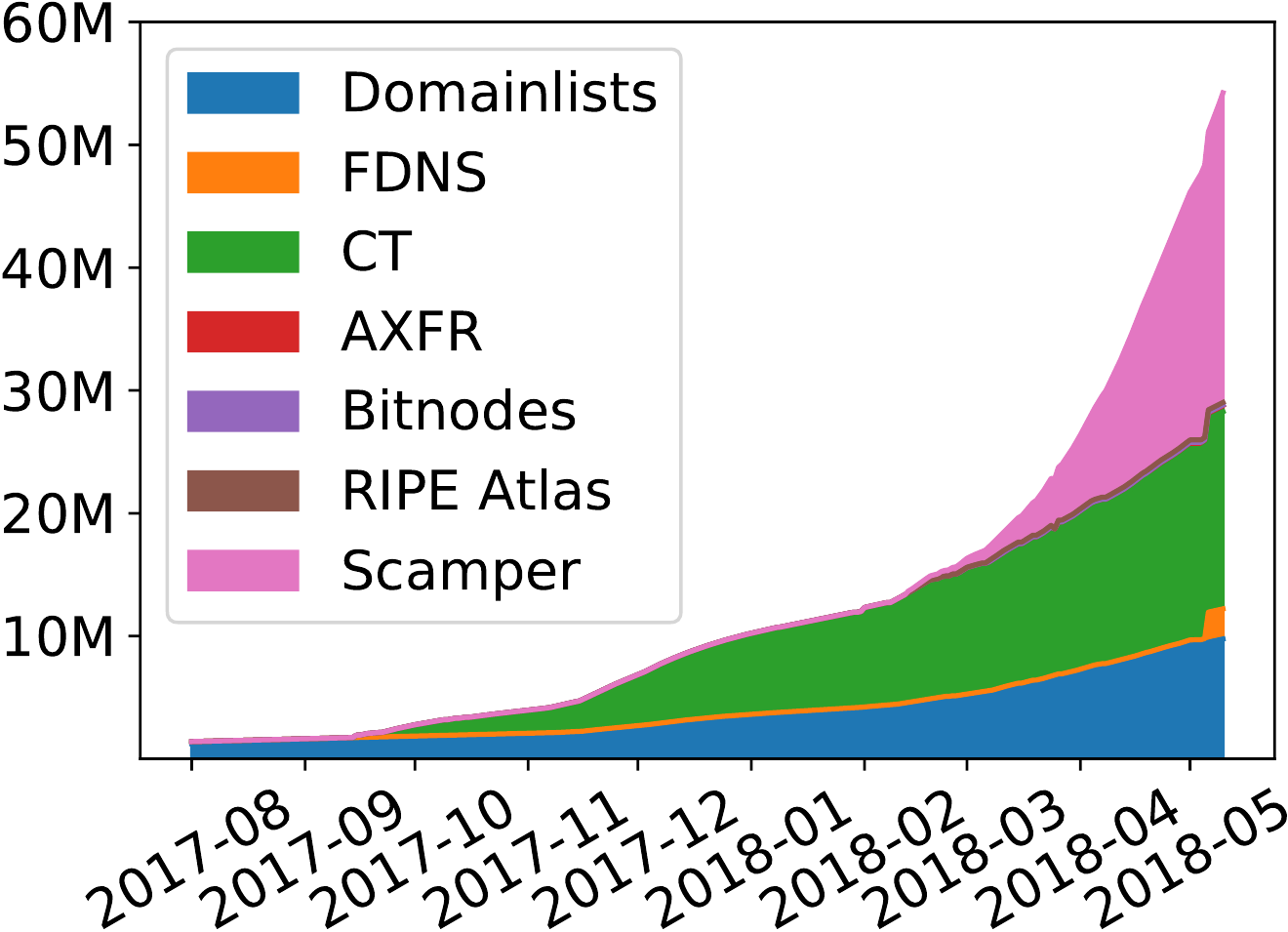}
		\caption{Cumulative runup of IPv6 addresses.}
		\label{fig:sources}
	\end{subfigure}%
    \hfill
	\begin{subfigure}{0.3\textwidth}%
    \includegraphics[width=\columnwidth]{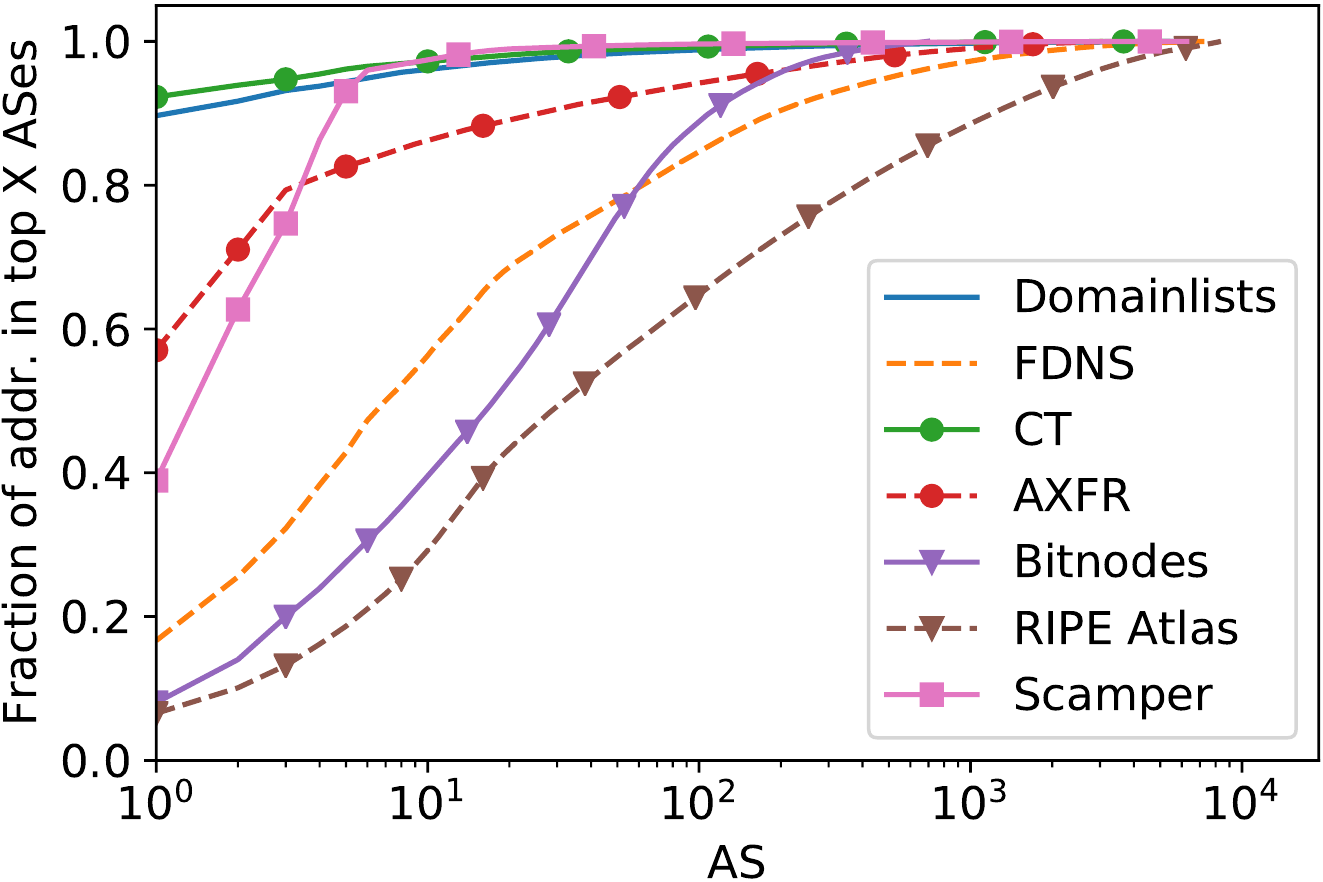}
    \caption{AS distribution for hitlist sources.}
    \label{fig:assources}
	\end{subfigure}%
    \hfill
    \begin{subfigure}{0.38\textwidth}%
    \href{https://ipv6hitlist.github.io/zesplot/#input}{%
    \includegraphics[width=\columnwidth]{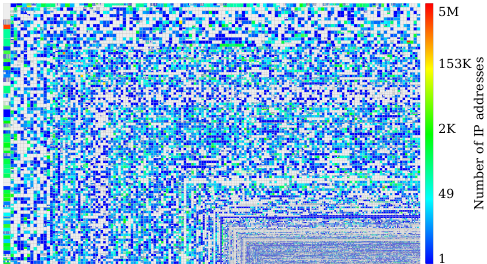}%
    }
    \caption{Hitlist addresses mapped to \sk{56} BGP prefixes.}
    \label{fig:zesplot:overview}
    \end{subfigure}%
    \caption{Hitlist sources runup, AS distribution as CDF, and zesplot.}
\end{figure*}

\textbf{Domain Lists: } Described in~\cite{amann2017mission, caastudy17,
pam18ctlog}, with a total of \sm{212} domains from various large zones, resolved
for AAAA records on a daily basis, yielding about \sm{9.8} unique IP
addresses.
This source also includes domains extracted from blacklists provided by Spamhaus \cite{spamhaus}, APWG \cite{apwg}, and Phishtank \cite{phishtank}, which leverage \sm{8.5}, \sk{376}, and \sk{170} domains, respectively.

\textbf{FDNS: } A comprehensive set of forward DNS (FDNS) ANY lookups performed by Rapid7~\cite{rapid7}, yielding \sm{2.5} unique addresses.

\textbf{CT: } DNS domains extracted from TLS certificates logged in Certificate Transparency (CT), and not already part of domain lists, which yields another \sm{16.2} addresses.

\textbf{AXFR and TLDR: } IPv6 addresses obtained from DNS zone transfers (AXFR)  from the TLDR project~\cite{tldr} and our own AXFR transfers. Obtained domain names are also resolved for AAAA records daily. This source yields \sm{0.5} unique IPv6 addresses.

\begin{figure*}[tb]
    \href{https://ipv6hitlist.github.io/eip/}{
\begin{subfigure}[b]{.47\linewidth}
	\includegraphics[width=\columnwidth]{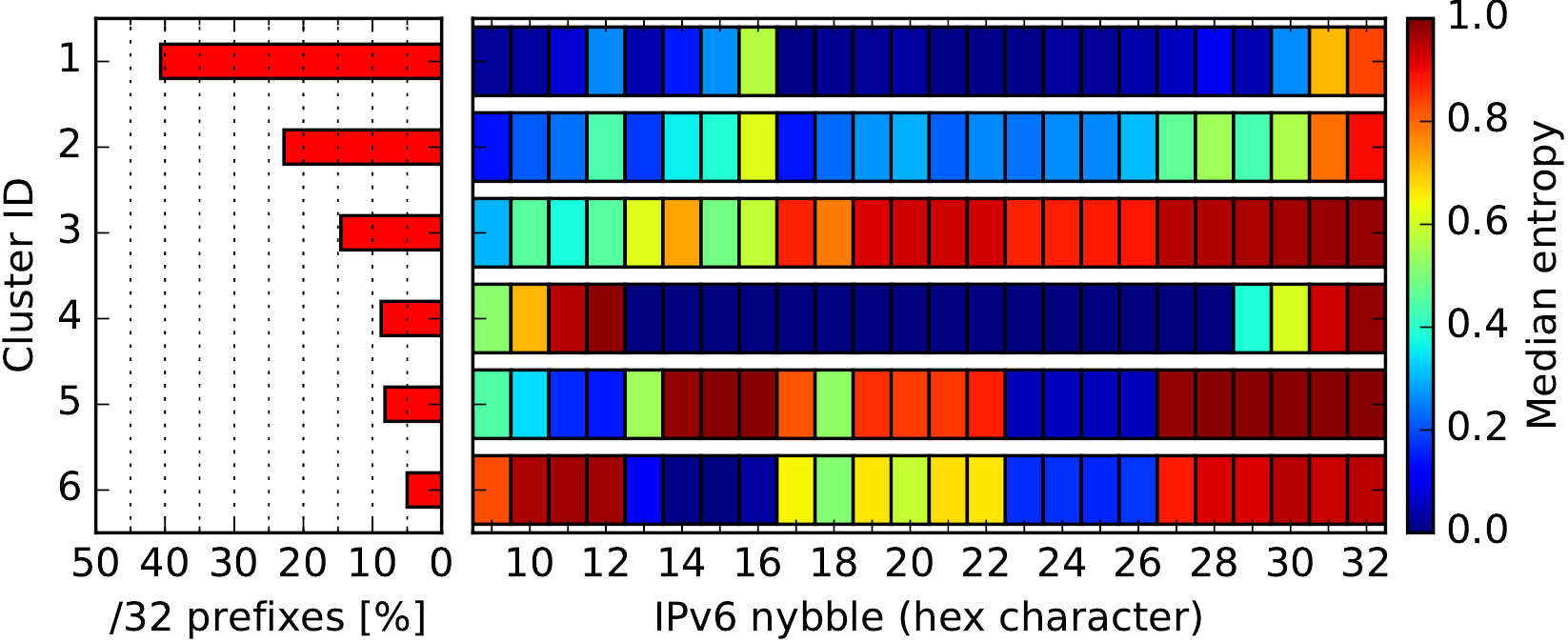}
	\caption{Fingerprints of full addresses, $\mathbf{F}^{9}_{32}$.}
	\label{fig:clusters-slash32}
\end{subfigure}
\hfill%
\begin{subfigure}[b]{.47\linewidth}
	\includegraphics[width=\columnwidth]{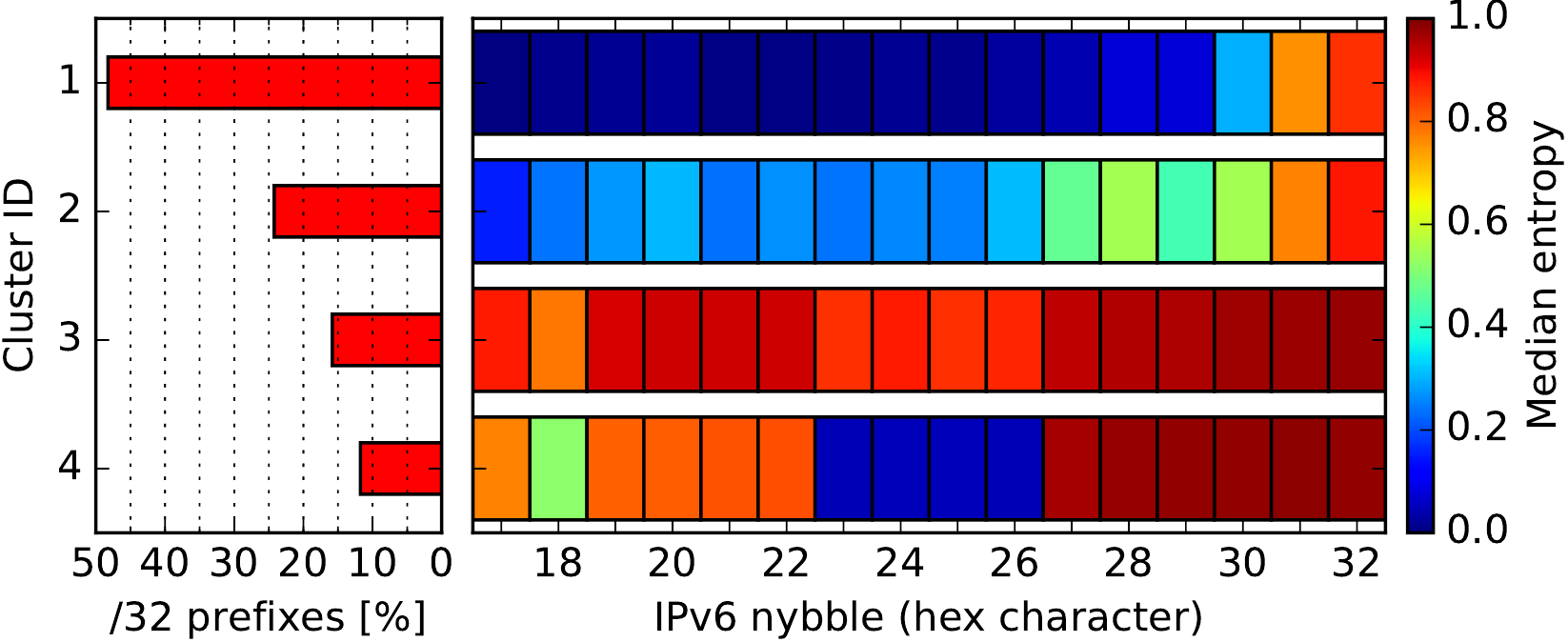}
	\caption{Fingerprints of IIDs, $\mathbf{F}^{17}_{32}$.}
	\label{fig:clusters-slash32-iid}
\end{subfigure}
}
\caption{/32 prefixes clustered using entropy fingerprints.}
\label{fig:clusters}
\end{figure*}

\textbf{Bitnodes: }To gather client IPv6 addresses, we use the Bitnodes API~\cite{bitnodes}, that provides current peers of the Bitcoin network.
Although this is the smallest source, contributing \sk{27} unique IPv6 addresses, we still find it valuable as it also adds client addresses.

\textbf{RIPE Atlas: }We extract all IPv6 addresses found in RIPE Atlas traceroutes, as well as all IPv6 addresses from RIPE's ipmap project~\cite{ipmap}, which adds another \sm{0.2} addresses. These are highly disjoint from previous sources, likely due to their nature as routers.

\textbf{Scamper: } Finally, we run traceroute measurements using scamper~\cite{luckie2010scamper} on all addresses from other sources, and extract router IP addresses learned from these measurements. This source shows a very strong growth characteristic, with \sm{25.9} unique IP addresses.

We accumulate all sources, \ie IP addresses will stay indefinitely in our scanning list. %
We may revisit this decision in the future, and remove IP addresses after a certain window of unresponsiveness.

\textbf{Address Runup: }\Cref{fig:sources} shows the cumulative runup of sources over time. 
First, we can see a strong growth of IPv6 addresses in all sources:
typically an increase by factors of 10--100 over the course of a year. 
Second, DNS-based sources---likely revealing server addresses---together with traceroute addresses obtained from scamper dominate the overall dataset.
As we find scamper's explosive growth peculiar, we conduct a closer investigation, which reveals that \sperc{90.7} of those IP addresses are SLAAC addresses, \ie marked by \texttt{ff:fe}. 
The vendor codes in MAC addresses gained from those routers indicate that they are typically home routers: \sperc{47.9} ZTE and \sperc{47.7} AVM (Fritzbox), followed by \sperc{1.2} Huawei with a long tail of 240 other vendors. 
This shows that our source includes mainly home routers and CPE equipment.
Depending on the type of study, it may be desirable to include or exclude these CPE devices.

\textbf{zesplot:}
To explore these large amounts of data, we present visualizations called
\textit{zesplots}. A zesplot visualizes IPv6 prefixes, each represented as a rectangle. It
does not show the entire IPv6 address space, only the prefixes provided as
input.  It uses a space-filling algorithm based on \textit{squarified
tree-maps}~\cite{bruls2000squarified}, extended with recursive properties. %
It starts out by filling a vertical row with rectangles, then a horizontal row,
then a vertical row again, and so on.
The prefixes are ordered based on \{prefix-size,ASN\}, so that
large prefixes are plotted in the top-left corner of the graph, and smaller
prefixes in the bottom-right corner, keeping similarly sized prefixes from the
same AS adjacent. Consequently, a prefix will be in the same spot in different zesplots, as long as the input prefixes are the same.
Axes have no meaning in a zesplot.
More-specific subprefixes
are plotted in the top half of that prefix's rectangle (a prefix with many more-specifics might result in a gray rectangle when zoomed-out).
A white rectangle means no addresses are present in that prefix.
For example, in ~\Cref{fig:zesplot:overview}, we find a bright red /19 in the
top-left, /32s around the center of the plot, and very specific prefixes like
/127s in the bottom-right part.
In addition to the list of prefixes,
the tool takes a list of addresses, and colors the prefix rectangles
based on the number of addresses that belong to that prefix.  With these
colors, one can quickly spot prefixes which are possibly over-represented in the
dataset, or verify certain assumptions (\eg larger prefixes cover
more addresses than smaller prefixes).
Depending on the dataset, an \textit{unsized} zesplot might provide extra
insights on \eg clusters of prefixes: in these plots, all rectangles are
equivalently sized, and the prefix size is only used for sorting.
We publish the zesplot tool~\cite{zesplot}
aiming at more
applications in measurement research.

\textbf{Input Distribution: }When evaluating the AS distribution for each source in \Cref{fig:assources}, we see stark differences, \eg for domainlists and CT only a handful of ASes make up a large fraction of addresses, compared to the more balanced RIPE Atlas source.
In addition, we analyze the distribution of \hitlist addresses to BGP prefixes using zesplot in \Cref{fig:zesplot:overview}.
We cover half of all announced BGP prefixes, but we find that some prefixes contain unusually large numbers of addresses, which we investigate in \Cref{sec:alias}.

\textbf{Comparison with DNSDB: } Our \hitlist covers \sperc{12.9} of IPv6 addresses, \sperc{69.4} of ASes, and \sperc{48.7} of BGP prefixes belonging to AAAA records stored in DNSDB~\cite{dnsdb}.
The majority of ``missing'' addresses belong to large CDN operators, which is probably due to DNSDB collecting passive DNS data globally, versus active probing from a few locations. Moreover, the majority of addresses in our hitlist are not in DNSDB, especially for infrastructure of ISP operators. We find our hitlist and DNSDB to be complementary.
We do not add DNSDB to our sources as it is not publicly available.

\textbf{Going Beyond: }
Besides the aforementioned daily scanned sources, we also conduct in-depth case studies on three disjunct input sources: 
newly learned IPv6 addresses using \sixgen and \eip in \Cref{sec:learning}, rDNS-walked IPv6 addresses in \Cref{sec:rdns}, and crowdsourced client IPv6 addresses in \Cref{sec:client}.

\section{Entropy Clustering} \label{sec:clustering:entropy}

\begin{figure*}
\begin{subfigure}{.475\linewidth}
    \href{https://ipv6hitlist.github.io/eip-scanning/}{%
	\includegraphics[width=\columnwidth]{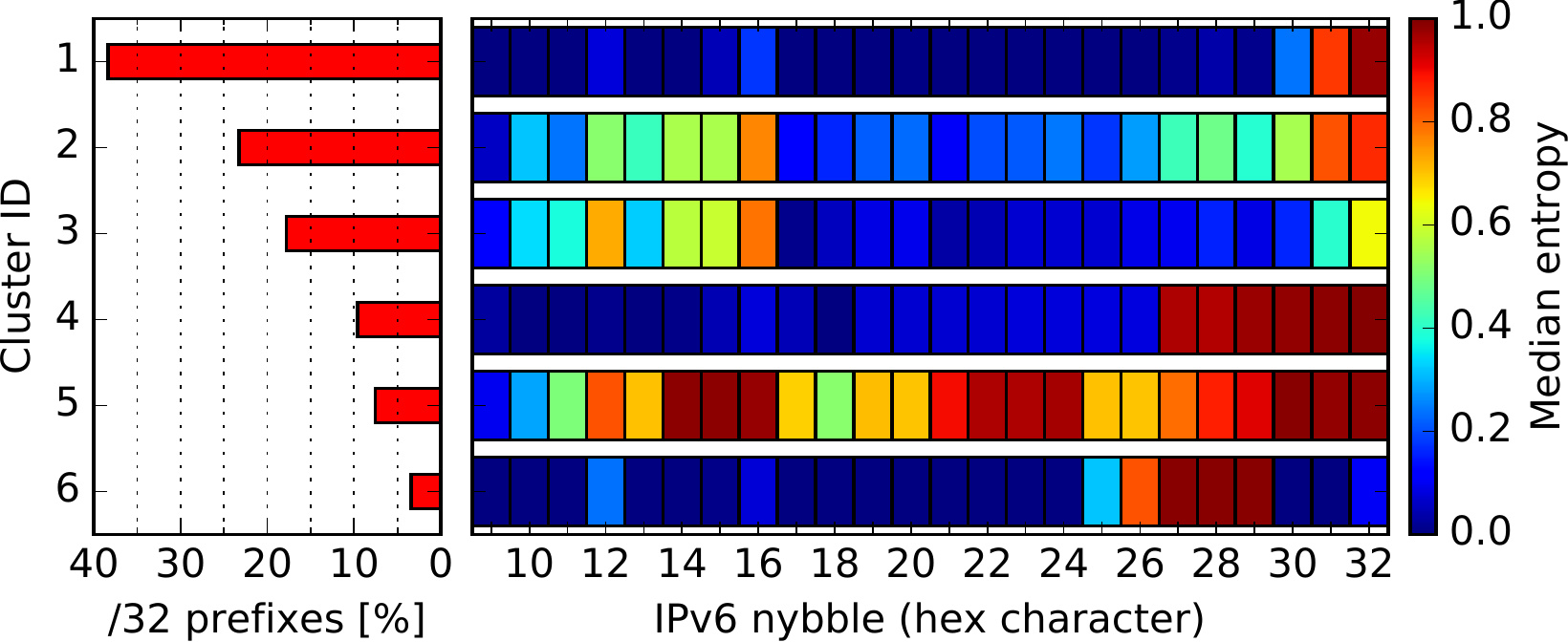}%
}
    \caption{/32 prefixes with addresses responding to UDP/53, clustered using entropy fingerprints of full addresses, $\mathbf{F}^{9}_{32}$.}
	\label{fig:clusters-dns}
\end{subfigure}
\hfill
\begin{subfigure}{.47\linewidth}
    \href{https://ipv6hitlist.github.io/zesplot/#clusters}{%
    \includegraphics[width=\columnwidth]{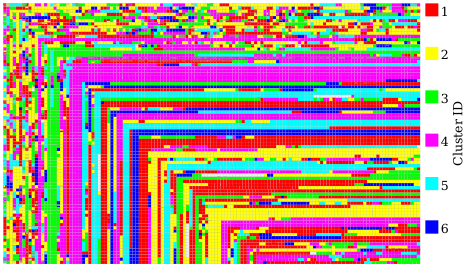}%
}
\caption{\sk{22} BGP prefixes colored using their $\mathbf{F}^{9}_{32}$ clusters.}
    \label{fig:zesplot:clusters-full-unsized}
\end{subfigure}
    \caption{Entropy clustering for DNS responsive hosts and cluster distribution in BGP prefixes.}
\end{figure*}

We introduce a method for grouping IPv6 networks by similar entropy in their addresses and evaluate it on our IPv6 hitlist.

Let $S$ be a set of IPv6 addresses in a particular network, \eg a /32 prefix, a
BGP prefix, or an AS. The set may be a random sample, but with at least 100 addresses.
We introduce the following notation:
\begin{align}
S =  &\; \lbrace \: \mathbf{A}_{1}^{}, \cdots, \mathbf{A}_{i}^{}, \cdots, \mathbf{A}_{n} \: \rbrace, \: n \geq 100 \\
\mathbf{A}_{i} = &\; ( \: x_{1}^{i}, \cdots, x_{j}^{i}, \cdots, x_{32}^{i} \: ), \\
x_{j}^{i} \in \Omega = &\; \lbrace  `0`, \cdots, `f` \rbrace,
\end{align}
where $\mathbf{A}_{i}$ is an address in that network: a sequence of 32 nybbles, \ie hex characters.
Let $X_{j}$ be a discrete random variable on $\Omega$ representing nybble $j$ across $S$, and have
an empirical probability mass function  $\hat{P}(X_{j})$. Next, we compute the normalized Shannon entropy of $X_j$, which we call an entropy fingerprint $\mathbf{F}^{a}_{b}$:
\begin{align}
\mathbf{F}^{a}_{b} = &\; ( \: H(X_{a}), \cdots, H(X_{j}), \cdots, H(X_{b}) \: ) \\
H(X_{j})  = &\; \frac{1}{4} \cdot -\sum_{\omega \in \Omega} \hat{P}(X_{j}=\omega) \cdot log \hat{P}(X_{j}=\omega).
\end{align}
where $a$ and $b$ are the first and the last considered nybbles, respectively. Note that if
$H(X_j)=0$, then nybble $j$ is constant; if $H(X_j)=1$, then all its values are equally
probable.

We repeat the above for each network, obtaining a dataset of fingerprints. Next, we run the k-means algorithm on the obtained dataset to find clusters of networks with similar fingerprints. We use the well-known elbow method to find the number of
clusters, $k$, plotting the sum of squared errors (SSE) for $k = \lbrace 1, \cdots, 20
\rbrace$:
\begin{align}
SSE(k) = \sum_{c \in C_k} \sum_{\mathbf{E} \in c} d^2(\mathbf{E}, \overline{\mathbf{c}})
\end{align}
where $C_k$ is the set of clusters obtained for given $k$, $\overline{\mathbf{c}}$ is the cluster
mean, and $d^2$ is the squared Euclidean distance. We select the $k$ for which we see an ``elbow''
in the plot, \ie the point where increasing $k$ does not yield a relatively large reduction in
$SSE$.

Finally, we summarize each cluster graphically with its median entropy on each nybble and
with its relative popularity.

\subsection{Results}

We present the results of entropy clustering on /32 prefixes in our hitlist in
\Cref{fig:clusters}, for two different fingerprint lengths. Each plot shows cluster popularity on the left-hand side, and the median entropy of each nybble on the right-hand side. Clusters are represented by rows, ordered top-down by their popularity.

Most notably, in \Cref{fig:clusters-slash32}, we identify just 6 clusters of full address fingerprints. The most popular cluster has entropy $\approx0$ on all nybbles but a few at the end of network and interface identifiers (IIDs). It is likely an artifact of a common practice to treat these parts of IPv6 addresses as simple counters. 
The second most popular pattern is similar, but uses more nybbles and introduces more structure. In cluster~3, we see prefixes with pseudo-random IIDs, manifested in high entropy, which makes predicting addresses in such networks impossible.
Finally, we see MAC-based IIDs in clusters~5 and~6, where nybbles 23-26 are likely just \texttt{ff:fe}.

In \Cref{fig:clusters-slash32-iid}, we focus on IIDs only, \ie we limit fingerprints to nybbles 17-32, and find just 4 clusters. Again, the majority of evaluated networks use IIDs as counters, as visible in clusters 1 and 2. This is expected and common, \eg for pools of servers, but shows potential for probabilistic scanning. We see the impact of SLAAC in clusters 3 and 4: pseudo-random and MAC-based, respectively.

Finally, we use entropy clustering throughout the paper to help with the interpretation of our results.
First, we try to better understand the IPv6 scanning results detailed in Section \ref{sec:probing}. In \Cref{fig:clusters-dns}, we present clusters with addresses that respond to a UDP/53 (DNS) scan.
We find that most clusters exhibit low entropy on all but a few nybbles. This phenomenon makes probabilistic scanning for IPv6 DNS servers easy, which we demonstrate in Section \ref{sec:learning}.

In \Cref{fig:zesplot:clusters-full-unsized}, we show how clusters are distributed to BGP prefixes.
The prefixes are ordered by their length and origin AS, but the box size is
static instead of based on the prefix length, which helps in pattern spotting.
We plot only prefixes belonging to ASes with more than 100 addresses.
With the order running from top-left (larger prefixes) to bottom-right (smaller prefixes), we find that the mix of clusters is more heterogeneous for larger prefixes. Within the smaller prefixes, we observe greater consistency. Most of the large, equally colored chunks are equally sized prefixes in a single AS. This hints at operators using the same addressing scheme, or deploying equivalent equipment or setups, in their prefixes.

\subsection{Discussion}

We highlight that entropy clustering is different from \eip~\cite{foremski2016entropy}. Although both algorithms use Shannon entropy, our method finds high-level patterns \emph{across} networks, whereas \eip finds low-level patterns \emph{within} a network. However, the methods are complementary: for instance, entropy clustering can help in spotting networks susceptible to probabilistic scanning, while \eip can generate a hitlist for each of them.

We also stress that entropy clustering is not confined to /32 prefixes, or to the fingerprint lengths we presented in the paper. Note that /32 prefixes are commonly the smallest blocks assigned to IPv6 networks \cite{arinv6,ripev6,afrinicv6,apnicv6,lacnicv6}. Using different entropy clustering parameters allows the different particularities of IPv6 deployment to be investigated.
We provide supplemental results obtained from clustering based on ASes, BGP prefixes, and other fingerprints online. Similarly, using other clustering tools will lead to different results. We present a baseline set of parameters and tools based on common practice, leaving other options for future work.

In summary, entropy clustering is a new technique that simplifies Internet studies. Instead of treating IPv6 as an opaque and expansive addressing space, we can visualize actual addressing patterns and their popularity in one picture. Our results have implications for scanning the IPv6 Internet. For instance, the clusters in \Cref{fig:clusters} suggest people in general use IPv6 addresses in a limited number of ways, and that many IPv6 address nybbles are easy to predict. On the other hand, cluster \emph{popularities} represent our hitlist---and to some extent, IPv6 hitlists in general---rather than the Internet.

\section{Aliased Prefix Detection}\label{sec:alias}\label{sec:probing:sub:aliasprefixdetection}
Aliased network prefixes, \ie prefixes under which each possible IP address replies to queries, were already found when conducting IPv4 measurements \cite{alt2014uncovering}.
For IPv6 measurements, however, aliased prefixes pose a much more significant challenge as they can easily contribute vast numbers of addresses that map to the same server, \eg through the \texttt{IP\_FREEBIND} option in Linux.
This feature is already in use by CDNs \cite{cffreebind}, and was identified as a challenge in previous works \cite{fiebig2017something,murdock2017target}. Aliased prefixes can artificially inflate the number of IP addresses within a \hitlist (\eg enumerating a /96 prefix can add $2^{32}$ addresses), and introduce significant bias into any studies using these \hitlists. 
Given this, we want to populate our \hitlist only with valuable addresses, \ie addresses belonging to different hosts and having balanced prefix and AS distributions.
This requires reliable detection and removal of aliased prefixes, for which we introduce a rigorous method in the following.

Similar to \cite{fiebig2017something,murdock2017target}, our method has its roots in the concept that a randomly selected IP address in the vast IPv6 space is unlikely to respond.
Thus, when probing randomly selected addresses, a prefix can be classified as aliased after a certain number of replies have been received.
Murdock \etal \cite{murdock2017target} send three probes each to three random addresses in every /96 prefix.
Upon receipt of replies from all three random addresses, the prefix is determined as aliased.
In the following, we describe how we improve efficiency and effectiveness of this approach in several ways.

Alias detection needs to fulfill two criteria to scale:
\one detection must be low-bandwidth, with a small number of packets required per network,
\two detection must function for end hosts, not only routers, which excludes many alias detection techniques.

\subsection{Multi-Level Aliased Prefix Detection}

For our daily scans, we perform multi-level aliased prefix detection (APD), \ie detection at different prefix lengths.
This is in contrast to previous works that use static prefix lengths, \eg ~/96.

To determine whether a prefix is aliased, we send 16 packets to pseudo-random addresses within the prefix, using TCP/80 and ICMPv6.
For each packet we enforce traversal of a subprefix with a different nybble.
For example, to check if \texttt{2001:db8:407:8000::/64} is aliased, we generate 1 pseudo-random address for each 4-bit subprefix, \texttt{2001:db8:407:8000:[0-f]000::/68}.
See \Cref{tab:fanout} for a visual explanation. 
Using this technique we ensure that \one probes are distributed evenly over more specific subprefixes and \two pseudo-random IP addresses, which are unlikely to respond, are targeted.
\begin{table}%
	\centering
    \caption{Example of IPv6 fan-out for multi-level aliased prefix detection. We generate one pseudo-random address in \texttt{2001:0db8:407:8000::/64} for each of the 16 subprefixes, \ie \texttt{2001:0db8:407:8000:[0-f]/68}.}
	\label{tab:fanout}
	\small
    \begin{tabular}{l}
        \toprule
        \texttt{2001:0db8:0407:8000::/64} \\
        \midrule
        \texttt{2001:0db8:0407:8000:\underline{0}151:2900:77e9:03a8} \\
        \texttt{2001:0db8:0407:8000:\underline{1}81c:4fcb:8ca8:7c64} \\
        \texttt{2001:0db8:0407:8000:\underline{2}3d1:5e8e:3453:8268} \\
        \multicolumn{1}{c}{\vdots} \\
        \texttt{2001:0db8:0407:8000:\underline{f}693:2443:915e:1d2e} \\
        \bottomrule
    \end{tabular}
\end{table}
For each probed prefix we count the number of responsive addresses. 
If we obtain responses from all 16 probed addresses, we label the prefix as aliased.

We run the aliased prefix detection on IPv6 addresses that are either BGP-announced or in our \hitlist.
The former source allows us to understand the aliased prefix phenomenon on a global scale, even for prefixes where we do not have any targets.
The latter source allows us to inspect our target prefixes more in-depth.

For BGP-based probing, we use each prefix as announced, without enumerating additional prefixes. 
For our \hitlist, we map the contained addresses to all prefixes from 64 to 124, in 4-bit steps.
We limit APD probing to prefixes with more than 100 targets for two reasons: First, efficiency, as APD probing requires 32 probes (16 for ICMPv6 and TCP/80, respectively). Second, impact, as prefixes with less than 100 probes can only distort our hitlist in a minor way.
We exempt /64 prefixes from this limitation so as to allow full analysis of all known /64 prefixes.
We use \sm{47.4} probes to guarantee complete coverage of all /64 prefixes and \sm{49.2} probes in total.

As we perform target-based APD at several prefix lengths, the following four cases may occur:
\begin{enumerate}%
	  \setlength{\itemsep}{0pt}
	    \setlength{\parskip}{0pt}
	    \setlength{\topsep}{0pt}	    
	    \vspace{-0.25em}
    \item Both more and less specific are aliased
    \item Both more and less specific are non-aliased
    \item More specific aliased, less specific non-aliased
    \item More specific non-aliased, less specific aliased
\vspace{-0.25em}
\end{enumerate}
The first two cases depict the ``regular'' aliased and non-aliased behaviors, respectively.
The third case is more interesting as we observe divergent results based on the prefix length that we query.
One example is a /96 prefix which is determined as being non-aliased, with only 9 out of 16 /100 subprefixes determined as aliased.
This case underlines the need for our fan-out pseudo-random aliased prefix detection.
Using purely random addresses, all 16 could by chance fall into the 9 aliased subprefixes, which would then lead to incorrectly labeling the entire /96 prefix as aliased.
The fourth case is an anomaly, since an aliased prefix should not have more specific non-aliased subprefixes.
One reason for this anomaly is packet loss for subprefix probes, incorrectly labeling the subprefix as non-aliased.

We analyze how common the fourth case is in our results and investigate the reasons.
On May 4, 2018, we detect only eight such cases at the prefix lengths /80, /116, and /120: %

The /80 prefix shows 3 to 5 out of the 16 possible responses over time.
The branches of responding probes differ between days, with no discernible pattern.
We suspect this prefix is behind a SYN proxy~\cite{rfc4987}, which is activated only after a certain threshold of connection attempts is reached.
Once active, the SYN proxy responds to every incoming TCP SYN, no matter the destination.

The /116 prefix consistently shows 15 out of 16 probes being answered on consecutive days, even though a less specific prefix was classified as aliased.
Moreover, the 15 probes answer with the same TCP options on consecutive days.
The non-responding probe is always on the \texttt{0x0} branch, so we believe the subprefix is handled differently and not by an aliased system.
In fact, comparing the paths of the different branches reveals that the \texttt{0x0} branch is answered by an address in a different prefix.
The DNS reverse pointer of this address hints at a peering router at DE-CIX in Frankfurt, Germany.
This /116 anomaly underlines the importance of the multi-level aliased prefix detection, since there are in fact small non-aliased subprefixes within aliased less specific prefixes.

The case of six neighboring /120 prefixes manifests less consistently than the previously described phenomenon.
The branches that lack responses change from day to day, as well as from prefix to prefix.
Subsequent manual measurements show that previously unresponsive branches become responsive.
The root cause is most likely ICMP rate limiting, which explains the seemingly random responding branches.
We try to counter packet and ICMP-rate limiting loss as explained in \Cref{sec:probing:sub:resilience}.

After the APD probing, we perform longest-prefix matching to determine whether a specific IPv6 address falls into an aliased prefix or not.
This ensures we use the result of the most closely covering prefix for each IPv6 address, which creates an accurate filter for aliased prefixes. %
If a target IP address falls into an aliased prefix, we remove it from that day's \zmapsix and \scamper scans.

\subsection{Loss Resilience}
\label{sec:probing:sub:resilience}

Packet loss might cause a false negative, \ie an aliased prefix being incorrectly labeled as non-aliased.
To increase resilience against packet loss, we apply \one cross-protocol response merging and \two a multi-day sliding window.

As we are probing all 16 target addresses on ICMPv6 and TCP/80, IP addresses may respond inconsistently. %
Our technique hinges on the fact that it is unlikely for a randomly chosen IP address to respond at all, so we treat an address as responsive even if it replies to only the ICMPv6 or the TCP/80 probe. %
While this greatly stabilizes our results, we still see high-loss networks, which would fail automatic detection, but could be manually confirmed as aliased.

To further tackle these, we introduce a sliding window over several past days, and require each IP address to have responded to any protocol in the past days. 
As prefixes may change their nature, we perform this step very carefully, and aim for a very short sliding window to react to such changes as quickly as possible.

To find an optimum, we compare the number of days in the sliding window to the number of prefixes that are unstable, \ie change the nature of stable and unstable over several days. 
We show the data in \Cref{tab:sw}, which confirms that with a sliding window of just 3 days, we can reduce the number of unstable prefixes by almost \sperc{80}, while 
only adding a small delay for prefixes that change their nature. 
With the final sliding window of 3 days, only 14 of the 909 aliased prefixes as of May 1, 2018 show an unstable nature. 

\begin{table}
	\caption{Impact of sliding window on unstable prefix count.}
	\vspace{-3mm}
	\label{tab:sw}
	\centering
	\small
	\begin{tabular}{lrrrrrr}
		\toprule
        Sliding window  & \num{0} & \num{1} & \num{2} & \num{3} & \num{4} & \num{5}   \\
		\midrule
        Unstable prefixes & \num{65} & \num{26} & \num{22} & \num{14} & \num{14} & \num{13}  \\
		\bottomrule
	\end{tabular}
\end{table}

\subsection{Impact of De-Aliasing}

We apply the aliased prefix detection (APD) filtering daily and analyze the impact on our \hitlist for May 11, 2018.
Before filtering, there are a total of \sm{55.1} IPv6 addresses on our hitlist.
After identifying aliased prefixes, \sm{29.4} targets (\sperc{53.4}) remain.
With the unfiltered hitlist we cover \num{10866} ASes and \num{25465} announced prefixes.
Removing aliased prefixes reduces our AS coverage by only 13 ASes, and lowers prefix coverage by \sperc{3.2} to \num{24648} prefixes.

\begin{figure}
    \centering
    \includegraphics[width=.95\columnwidth]{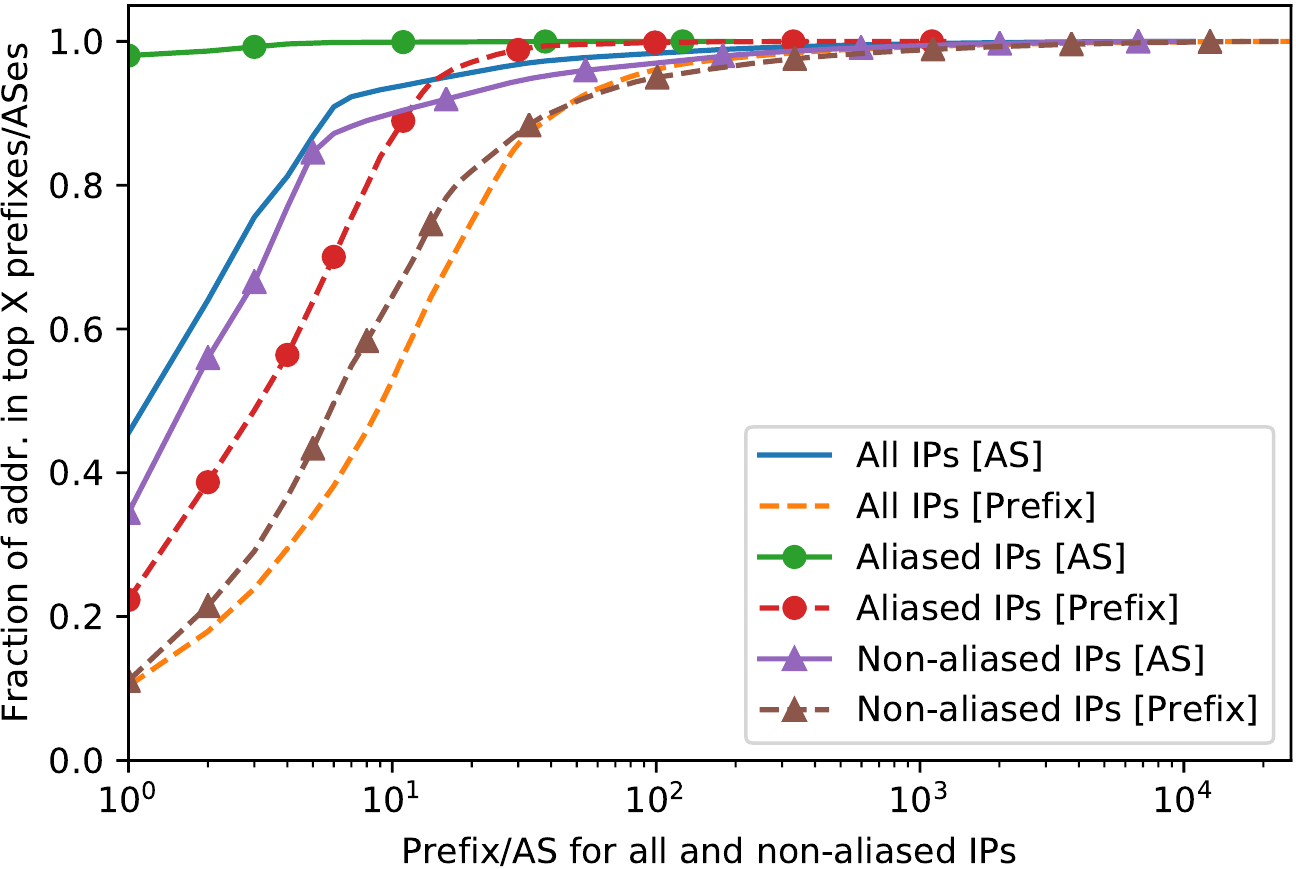}
    \caption{Prefix and AS distribution for aliased, non-aliased, and all hitlist addresses.}
    \label{fig:asdistalias}
\end{figure}

We show the AS and prefix distribution for aliased, non-aliased, and all IPv6 addresses in \Cref{fig:asdistalias}.
By comparing AS distributions we find aliased prefixes as heavily centered on a single AS (Amazon). %
In consequence, the AS distribution for non-aliased prefixes is flatter than the population as a whole. %
The picture changes for prefix distributions:
targets in non-aliased prefixes are now slightly more top-heavy compared to the general population.
One of the reasons is that the vast majority of aliased IP addresses are within 189 /48 prefixes announced by Amazon, which are the shortest detected aliased prefixes.
This results in shifting down the prefix distribution of the general population.

\begin{figure}[!h]
    \begin{subfigure}[]{\columnwidth}
    \href{https://ipv6hitlist.github.io/zesplot/#apd}{%
    \includegraphics[width=\columnwidth,keepaspectratio]{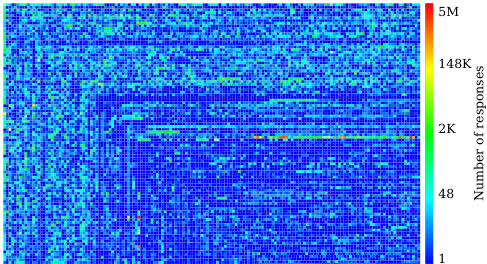}%
}
        \caption{\sk{16} prefixes without aliased prefix detection.}
        \label{fig:zesplot:without-ad}
    \end{subfigure}
    \begin{subfigure}[]{\columnwidth}
    \href{https://ipv6hitlist.github.io/zesplot/#apd2}{%
        \includegraphics[width=\columnwidth,keepaspectratio]{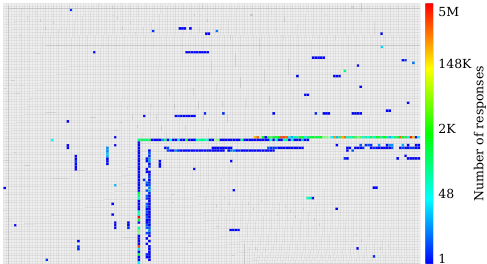}%
}
        \caption{461 (\sperc{3.0} of \sk{16}) detected aliased prefixes.}
        \label{fig:zesplot:with-ad}
    \end{subfigure}
    \caption{Responses to ICMP Echo requests.}
\end{figure}

To further visualize the effect of filtering aliased prefixes, we
compare~\Cref{fig:zesplot:without-ad} with~\Cref{fig:zesplot:with-ad}.
In these figures, equally sized boxes are plotted for every prefix, ordered by prefix length and ASN.
We find that aliasing barely occurs in the shortest prefixes, plotted in the top left corner of the plot.
Towards the lower right end---\ie the longer prefixes---we find some groups of aliased /32 prefixes, but most eye-catching are the groups forming the big ``hook''.
These are all /48 prefixes, with the majority belonging to Amazon (the ``outer'' hook) and Incapsula (the ``inner'' hook).
We also see that filtering these prefixes is effective:
the brightly colored Amazon aliased prefixes comprise a large share of the input set.

\subsection{Fingerprinting Aliased Prefixes}
The reason for not scanning aliased prefixes is that each contained IP address is assumed to belong to the same host, and consequently to display the same properties. 
We investigate this assumption by deploying fingerprinting techniques in our probes. 
Employing a ZMap module that supports the TCP Options header~\cite{zmaptcpopts}, 
we send a set of commonly supported fingerprinting options (\texttt{MSS-SACK-TS-WS}), setting MSS and WS to 1 to trigger differing replies~\cite{siblings}. 

Fingerprinting is a fuzzy and challenging technique, which is why we carefully evaluate results and consider them indicative rather than conclusive.
We consider this as a case study for validation, and do not feed the results back to scanning. 

Below, we investigate replies from \num{20692} /64 prefixes classified as aliased, for which all of our 16 APD probes to TCP/80 succeeded on May 11, 2018. We first analyze TTL values of response packets.
Previous work found that TTL values cannot be expected to be constant per prefix or even IP address~\cite{ttl, backes2016feasibility}.
TTL inconsistencies can stem from routing changes, TTL-manipulating middleboxes, or other on-path effects.
We quickly re-appraise this sentiment with our data, and can confirm that \num{5970} of our \num{20692} prefixes offer inconsistent TTL values. 
We hence replace the raw TTL metric with the likely initial TTL (iTTL) value chosen by a host.

\textbf{iTTL:} Rounding the TTL value up to the next power of 2 results in the iTTL value~\cite{mukaddam2014ip,jin2003hop,backes2016feasibility}. 
Using iTTL, we find only 6 prefixes with inconsistent behavior, all caused by 22 IP addresses responding with differing iTTL values to our 2 consecutive probes. 
These addresses belong to only 2 /48 prefixes, announced by 2 not commonly known ASes. 
As the iTTL can be only one of four values (32, 64, 128, or 255), we use differing iTTL as a negative indicator for APD: many different iTTL values suggest a non-aliased prefix.
With only \sperc{0.03} of inconsistent prefixes, we remain confident in our APD filtering. 

\textbf{Optionstext:} We next evaluate a metric used by~\cite{siblings, beverly2015server} that translates TCP options into a string, which preserves the order of options and padding bytes, but not the option values. 
For example, the string \texttt{MSS-SACK-TS-N-WS} would represent a packet that set the Maximum Segment Size, Selective ACK, Timestamps, a padding byte, and Window Scale options. 
Although we found \sperc{99.5} of responsive hosts to choose that set of options, 
we identify 104 prefixes that return differing sets of TCP options, an unlikely behavior for a prefix aliased to the same machine. 

\textbf{WScale and WSize:} The next metric is TCP window size and TCP window scale option.
These options are also not necessarily expected to stay constant, as changes in host state can lead to advertising varying window sizes. 
However, a consistent window across all IP addresses in a prefix again raises our confidence in our technique.
We find \num{1068} prefixes with inconsistent TCP window sizes, and \num{105} prefixes with inconsistent window scale options: a sizable number, but amounting to only $\approx\sperc{5}$ of our aliased prefixes. 

\textbf{MSS: }
Like iTTL, the TCP Maximum Segment Size is used as a negative indicator.
Inconsistent MSS values are determined for 1030 prefixes. 
This behavior is, as for previous metrics, typically caused by individual IP addresses sending differing  MSS sizes. %

\textbf{Timestamps:} Finally, we evaluate TCP Timestamps, as suggested by~\cite{siblings, beverly2015server}.
Although TCP Timestamps offer highly discriminative features, they cannot reliably identify a prefix as non-aliased with only 2 probes per host, due to the variety of possible behaviors, \eg randomized start values. They can, however, strengthen our confidence that an aliased prefix is behaving consistently.
We hence run the following checks:
\one whether all hosts send the same (or missing) timestamps,
\two whether timestamps are monotonic for the whole prefix, and
\three whether the receive timestamp and remote TCP timestamp have a regression coefficient $R^2>0.8$. 
This tests for a global linear counter, strongly hinting that the queried IP addresses belong to the same machine~\cite{siblings, beverly2015server}.
If any of these three tests succeeds, we consider a prefix to offer consistent behavior, which is a strong indicator for aliasing.

We find \num{13202} of the \num{20692} prefixes to exhibit a consistent behavior. 
Given that all Linux machines since kernel 4.10 would fail our tests as they randomize initial timestamps per \texttt{<SRC-IP, DST-IP>} tuple~\cite{siblings}, we consider this to be a quite high indicator that our APD probing does indeed find aliased IP addresses.
Note that due to many valid scenarios, a failed timestamping test does not make a prefix inconsistent, but is simply indecisive as to whether a prefix may or may not be aliased.

\begin{table}[tb]
	\centering
	\caption{Fingerprinting \sk{20.7} aliased prefixes: inconsistent prefixes per test, in total, and total consistent.}
	\begin{tabular}{lrrr}
		\toprule
		Test       & Incs. & $\Sigma$\,Incs. &  $\Sigma$\,Cons.  \\
		\midrule
        iTTL       &            \num{6} &                   \num{6}   &  \num{20686} \\
        Optionstext    &          \num{104} &                 \num{110}  & \num{20581}  \\
        WScale     &          \num{105} &                  \num{215} &  \num{19515} \\
        MSS    &         \num{1030} &          \num{1175}         &  \num{19513} \\
        WSize &         \num{1068} &           \num{1186}        &  \num{19506}  \\
		Timestamps      &          n/a\textsuperscript{1}    & n/a\textsuperscript{1} & \num{13202} \\
		\bottomrule
		\multicolumn{4}{l}{\textsuperscript{1}{\small A failed timestamping test does not indicate}} \\
		\multicolumn{4}{l}{\small ~an inconsistent prefix, but an indecisive one.}
	\end{tabular}
	\label{tab:fpalias}
\end{table}

\begin{table}[tb]
 	\centering
	\caption{Validation: aliased prefixes are considerably less inconsistent and pass many more consistency checks, including timestamping.}
	\begin{tabular}{lrrrr}
		\toprule
		Scan type       & Incons. & Cons. & Indec. \\
        \midrule
        Non-aliased prefixes &             \sperc{50.4}         &  \sperc{23.8} & \sperc{25.8} \\
        Aliased prefixes    &   \sperc{5.1}        &      \sperc{63.8}          & \sperc{31.1}  \\
		\bottomrule
	\end{tabular}
	\label{tab:fpalias_validation}
\end{table}

\Cref{tab:fpalias} shows statistics for all performed consistency tests.
Excluding TCP Timestamps, all tests combined find only \num{1186} inconsistent prefixes.
Interestingly, $>\sperc{90}$ of those are caused by hosts showing surprisingly inconsistent behavior to our 2 probe packets. 
For example, we found 22 hosts responding with distinctively different iTTL values (64 vs. 255) in direct order. 
We also see hosts responding with different sets of TCP options or option values. 
Many of these are hosted at CDNs, and might be TCP-level proxies to other services, which could explain time-variant fingerprints. 

\textbf{Validation: }
To verify our methodology, we run the same tests on prefixes considered \emph{non}-aliased after several days of APD. %
For direct comparison, we only choose \num{2940} /64 prefixes with $\geq$16 responding IP addresses.
As shown in \Cref{tab:fpalias_validation}, we find \num{1481} (\sperc{50.4}) to fail at least one of our tests, a considerably higher share than the \sperc{5.1} among aliased prefixes. 
Additionally, we find only \num{699} prefixes (\sperc{23}) to pass our high-confidence test of consistent timestamping behavior, compared to \sperc{63.8} in aliased prefixes.

Looking deeper into the 699 allegedly non-aliased prefixes that pass our high-confidence test, we find 99 where the whole prefix sends the same TCP timestamp, 509 with strictly monotonic timestamp behavior, and 91 passing our $R^2$ test. 
As same and strictly monotonic TCP timestamps are unlikely to happen by chance, we investigate why those prefixes were not detected as aliased.
We find two root causes which can cause this scenario: 
\one prefixes may be aliased at subprefixes longer than /64, which we only probe if more than 100 IP addresses are responsive in that prefix---only 167 of the 699 prefixes qualified for probing longer subprefixes than /64. %
\two~prefixes may contain many IPv6 addresses specifically bound to one machine, but without binding the full prefix to the machine. This would cause a set of our random probes to fail, and hence the prefix would be detected as not aliased. We consider the remaining 167 prefixes to fall in this category. 

From this deep-dive, we conclude that those 699 prefixes that pass our timestamping test are indeed highly likely to have all their active IP addresses bound to the same machine, but without representing $>$100 IP addresses (532 prefixes), or without binding the full prefix (167 prefixes). Therefore, our APD detection algorithm does not detect these as aliased. 
This discrepancy stems from the facts that \one our APD, by not probing low-density prefixes, may give some false negatives, and \two the ``same machine'' test in our validation also holds true for prefixes with many IPv6 addresses specifically bound to one host, while our APD aims to find prefixes where \textit{all} IP addresses are bound to the same host. 

As we consider this validation step merely an informative case study, we do not consider these problematic.

In sum, our validation step shows that 
\one our tests are discriminatory and
\two aliased prefixes offer far less diverse configurations, with many more cases believed to be the same machine.

\subsection{Comparison to Murdock et al.'s Approach}

In order to assess our APD approach we quantitatively compare it to Murdock \etal's \cite{murdock2017target}.
As Murdock \etal perform alias detection on a best-effort basis by probing addresses in prefixes with a static prefix length of /96, we expect our approach to find more aliased prefixes.
This is in fact the case as we find \sk{992.6} \hitlist addresses residing in aliased prefixes which are not detected by Murdock \etal
On the other hand, Murdock \etal's approach classifies only \sk{1.4} \hitlist addresses as aliased which we deem non-aliased.

Additionally, we compare the bandwidth requirements of our APD approach to Murdock \etal's.
As our approach works on multiple prefix levels where at least 100 targets are present, we focus on the most likely aliased prefixes.
Consequently, we send probes to a total of \sm{50.1} IPv6 addresses to determine aliased prefixes in our hitlist.
Using Murdock \etal's static /96 prefixes, more than twice as many IPv6 addresses (\sm{113.8}) are probed.

To summarize, our approach finds \sk{992.6} more hitlist addresses in aliased prefixes compared to Murdock \etal's approach and at the same time probes less than half the number of IP addresses.
\section{Address Probing} \label{sec:probing}

We generate IPv6 targets and probe these targets' responsiveness each day.
First, we collect addresses from our hitlist sources.
Second, we preprocess, merge, and shuffle these addresses in order to prepare them as input for scanning.
Third, we perform aliased prefix detection to eliminate targets in aliased prefixes.
Fourth, we traceroute all known addresses using \scamper \cite{luckie2010scamper} to learn additional router addresses.
Fifth, we use \zmapsix~\cite{zmapv6} to conduct responsiveness measurements on all targets.
We send probes on ICMP, TCP/80, TCP/443, UDP/53, and UDP/443 to cover the most common services \cite{Gasser2016a}.
We repeat this process each day to allow for longitudinal responsiveness analysis.

\subsection{Responsive Addresses}

We first evaluate responsive addresses based on their corresponding BGP prefix.

\begin{figure}[!h]
    \href{https://ipv6hitlist.github.io/zesplot/#responses}{%
    \includegraphics[width=\columnwidth,keepaspectratio]{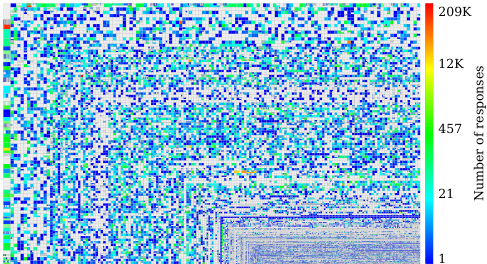}
}
    \caption{All \sk{56} BGP prefixes, colored based on the number of responses to ICMP Echo requests on May 11, 2018.}
    \label{fig:zesplot:probing-responses}
\end{figure}

Overall, our hitlist contains \sm{1.9} responsive IPv6 addresses, spread over \num{21647} BGP prefixes covering \num{9968} different ASes.

\Cref{fig:zesplot:probing-responses} shows non-aliased ICMP-responsive addresses per BGP prefix.
We see that most prefixes are covered with dozens to hundreds of responsive targets, whereas a few prefixes contribute \sk{12} or more responsive addresses.
The plot is, in terms of colors, strikingly similar to the input set visualized in~\Cref{fig:zesplot:overview} (note however that the range of the scale in the response plot is smaller).
This tells us that for most of the prefixes in our input set, we indeed see responses, and only few return no responses at all.
A possible explanation for prefixes with a sizable number of addresses in the input set ending up blank in the response plot is the dropping of ICMP echo requests at a border router.

\subsection{Cross-protocol Responsiveness}

We analyze the cross-protocol responsiveness of our probes, to understand what kind of IPv6 hosts are responding to our probes.
In \Cref{fig:protoheatmap} we show the conditional probability of responsiveness between protocols, \ie if protocol X is responding, how likely is it that protocol Y will respond.
We compare our findings for IPv6 to Bano \etal \cite{Bano:2018:SIL:3213232.3213234} who performed a similar analysis for IPv4.

\begin{figure}
    \centering
    \includegraphics[width=\columnwidth]{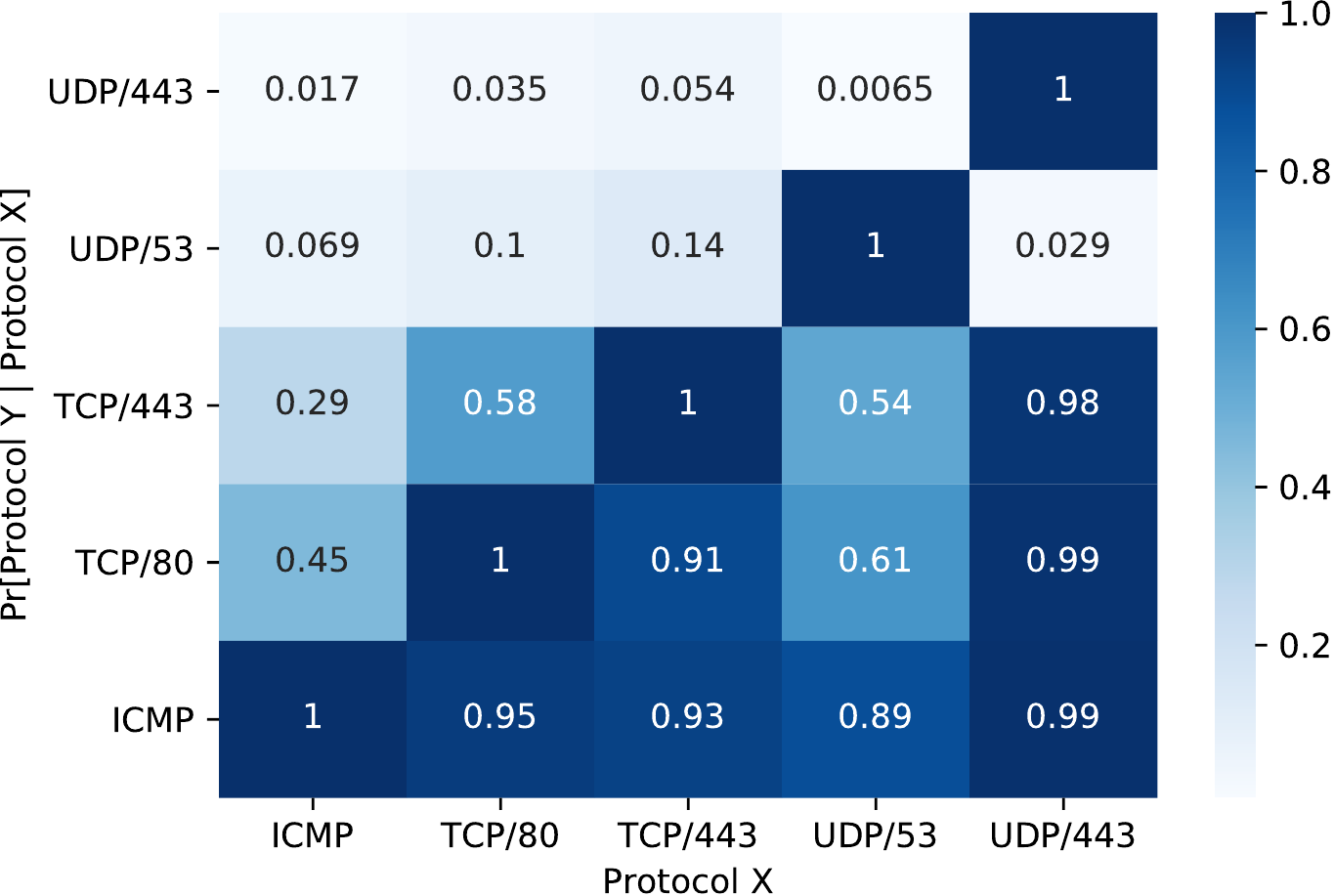}
    \caption{Conditional probability of responsiveness between services.}
    \label{fig:protoheatmap}
\end{figure}

We find that if an IPv6 address responds to any of the probes, there is at least a \sperc{89} chance of the same IP address also responding to ICMPv6.
The ICMP correlation in IPv6 is higher compared to IPv4, where we see values as low as \sperc{73}.
Since ICMPv6 is an integral part of IPv6, it should not be simply blocked in firewalls \cite{rfc4890}, which makes it more likely that hosts are responding to ICMPv6 compared to its IPv4 counterpart.

Additionally, we see correlations between UDP/443, TCP/443, and TCP/80.
More specifically, if an address is responsive to QUIC (UDP/443), it has a likelihood of \sperc{98} to be also providing HTTPS and HTTP services.
HTTPS servers are \sperc{91} likely to provide an HTTP service as well, \eg to offer a forwarding service to the secure version of a web page.
Note that the reverse correlation (HTTP $\rightarrow$ HTTPS $\rightarrow$ QUIC) is far less pronounced.
Compared to the HTTPS $\rightarrow$ HTTP correlation of \sperc{91} in IPv6, we see only a \sperc{72} correlation in IPv4 \cite{Bano:2018:SIL:3213232.3213234}.

Analyzing DNS (UDP/53) correlation shows mostly similar results in IPv6 as in IPv4.
One exception is the lower correlation to HTTPS in IPv6 (\sperc{54}) compared to IPv4 (\sperc{78}).

\subsection{Longitudinal Responsiveness}

To analyze address responsiveness over time, we probe an address continuously even if it disappears from our hitlist's daily input sources.
We evaluate longitudinal responsiveness over two weeks as depicted in \Cref{fig:rrheatmap}.
As a baseline for each source we take all responsive addresses on the first day.

\begin{figure}
    \centering
    \href{https://ipv6hitlist.github.io/probing/}{%
    \includegraphics[width=\columnwidth]{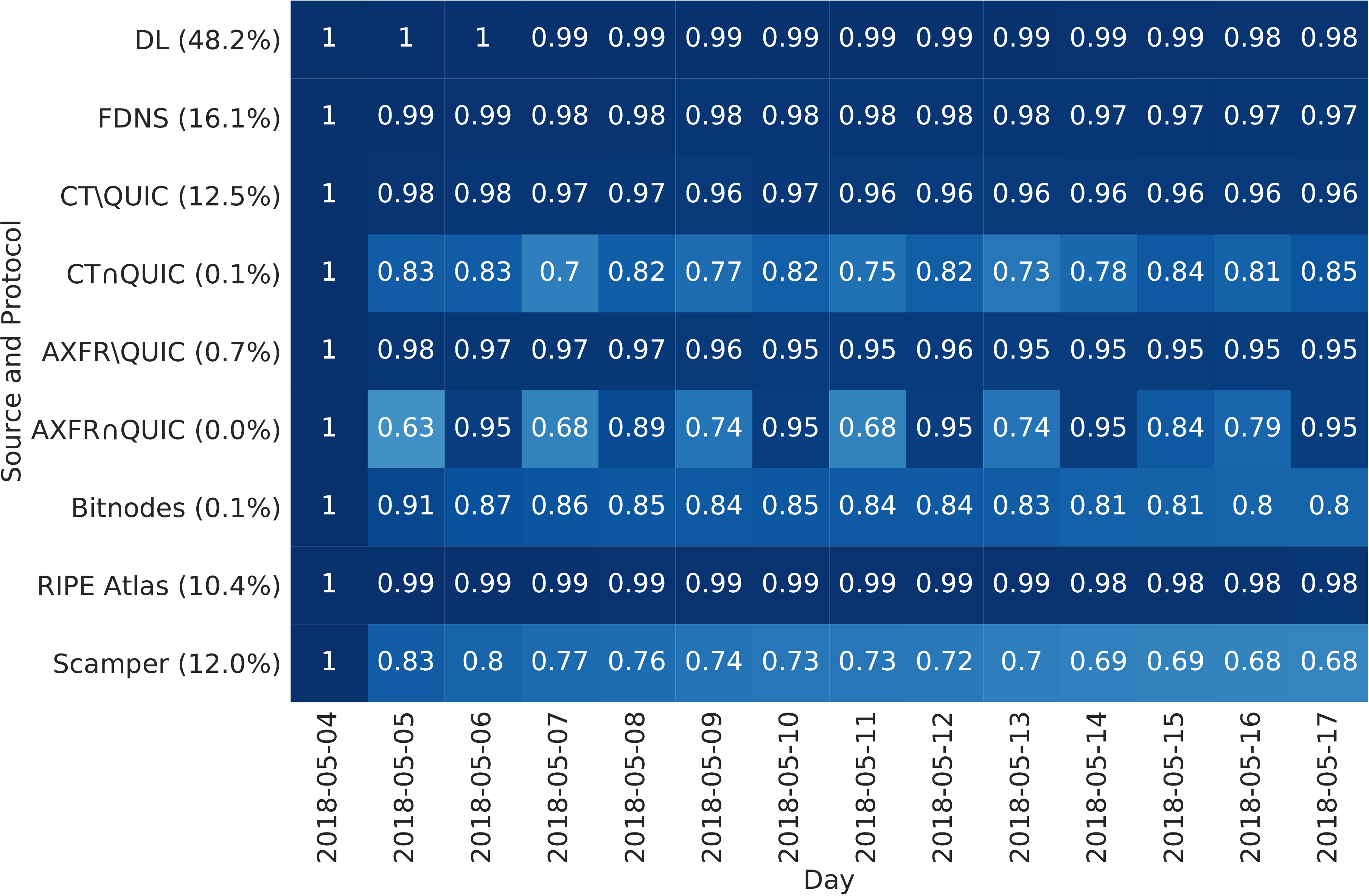}}
    \caption{Responsiveness over time, split up by hitlist source and, for special cases, also probed protocol.}
    \label{fig:rrheatmap}
\end{figure}

We find that IPv6 addresses from domain lists (DL), FDNS, and RIPE Atlas answer quite consistently over the 14 day period, with all three sources losing only a few percentages of addresses.
CT and AXFR sources overall reach a similarly stable response rate; their QUIC response rates, however, fluctuate more heavily and are therefore depicted separately.
We investigate this phenomenon and find that more than \sperc{80} of fluctuating addresses are located in two prefixes: Akamai and HDNet.
We suspect that these companies are testing the deployment of QUIC on some of their systems or that our measurements are caught by a rate limiting mechanism, resulting in flaky response behavior.
Moreover, sources which include clients or CPE devices such as Bitnodes and Scamper lose \sperc{20} and \sperc{32} of the responding hosts, respectively.
\section{Learning New Addresses} \label{sec:learning}

In addition to acquiring IPv6 addresses through domain names and other sources, we can also detect addressing schemes, and leverage those patterns to learn previously unknown addresses.

\subsection{Methodology}

To generate previously unknown addresses, we feed our hitlist into a re-implementation of \eip \cite{foremski2016entropy,entropy-ip,eip-generator}, and a pre-release version of \sixgen \cite{murdock2017target}.
For this work, we improve the address generator of \eip by walking the Bayesian network model exhaustively instead of randomly.
The improved generator lets us focus on more probable IPv6 addresses, under a constrained scanning budget.

First, we use all addresses in non-aliased prefixes to build a seed address list.
Excluding aliased prefixes avoids generating addresses in prefixes where all addresses are responsive, and thus artificially distorting the response rate.
Second, we split the seed address list based on ASes, as we assume similar addressing patterns within the same AS.
We limit the eligible ASes to those with at least 100 IPv6 addresses to increase the probability of \sixgen and \eip identifying patterns.
Third, we take a random sample of at most \sk{100} IPv6 addresses per AS to use as input for \sixgen and \eip.
The capped random sample ensures that we provide a balanced input for each AS.
Fourth, we run \eip and \sixgen with the capped random sample as input to generate \sm{1} addresses for each AS separately.
Fifth, we again take a random sample of at most \sk{100} of all generated addresses per AS for \sixgen and \eip, respectively.
The capped random sample ensures that ASes with more generated addresses are not overrepresented.
Sixth, we perform active measurements to assess the value of generated addresses.

\subsection{Learned Addresses}

\eip generates \sm{118} addresses.
Of those, \sm{116} are routable new addresses not yet in our hitlist.
\sixgen produces slightly more addresses, \sm{129}, of which \sm{124} are new and routable.
In total, we learn \sm{239} new unique addresses.
Interestingly, there is very little overlap between \sixgen's and \eip's generated addresses: only \sk{675} addresses are produced by both tools, which equals to \sperc{0.2} of all generated addresses.

\subsection{Responsiveness of Learned Addresses}

We probe the responsiveness of all \sm{239} learned addresses on ICMP, TCP/80, TCP/443, UDP/53, and UDP/443.
\sk{785} IPv6 addresses respond to our probes, which corresponds to a response rate of \sperc{0.3}.
This low response rate underlines the challenges of finding new responsive addresses through learning-based approaches.

Comparing the responsiveness of addresses generated by \sixgen to \eip, we find that \sixgen is able to find almost twice as many responsive addresses: \sk{489} vs. \sk{278}.
Our response rate for \sixgen is a lower bound: due to \sixgen's design, choosing the top generated addresses instead of random sampling would likely yield an even higher response rate.

In addition, both \eip and \sixgen found the same \sk{17} responsive addresses.
The response rate of overlapping addresses generated by both tools is therefore \sperc{2.5}, which is an order of magnitude higher than the general learned address population's \sperc{0.3}.
This demonstrates that \eip and \sixgen find complementing sets of responsive IPv6 addresses, with a small overlap of targets that are more likely to respond.
Thus, it is meaningful to run multiple address generation tools even on the same set of input addresses.

\begin{table}[!ht]
    \centering
    \caption{Top 5 responsive protocol combinations for \sixgen and \eip.}
    \label{tab:eip6genproto}
    \resizebox{\columnwidth}{!}{
    \begin{tabular}{cccccrr}
       \toprule
       ICMP & TCP/80 & TCP/443 & UDP/53 & UDP/443 & \sixgen & \eip \\
       \midrule
 \cmark &  \xmark &  \xmark &  \xmark &  \xmark &  \sperc{66.8} &  \sperc{41.1} \\
 \cmark &  \cmark &  \cmark &  \xmark &  \xmark &   \sperc{9.2} &  \sperc{12.3} \\
 \xmark &  \xmark &  \xmark &  \cmark &  \xmark &   \sperc{7.3} &  \sperc{23.1} \\
 \cmark &  \cmark &  \xmark &  \xmark &  \xmark &   \sperc{4.9} &   \sperc{3.4} \\
 \cmark &  \cmark &  \cmark &  \xmark &  \cmark &   \sperc{3.2} &   \sperc{6.1} \\
       \bottomrule
    \end{tabular}
    }
\end{table}

When analyzing the top 5 protocols for responsive learned addresses in \Cref{tab:eip6genproto}, we find particular differences between \sixgen and \eip.
Two thirds of \sixgen responsive addresses answer to ICMP only, which is the case for only four out of ten \eip responsive addresses.
On the other hand, \eip responsive hosts are three times more likely to be DNS servers (UDP/53).
Moreover, \sixgen responsive hosts are half as likely to be QUIC-enabled web servers (ICMP, TCP/80, TCP/443, and UDP/443) compared to \eip.
This shows that \sixgen and \eip not only discover mostly non-overlapping addresses, but also different types of \emph{populations} of responsive hosts.

\begin{figure}[tb]
    \includegraphics[width=\columnwidth]{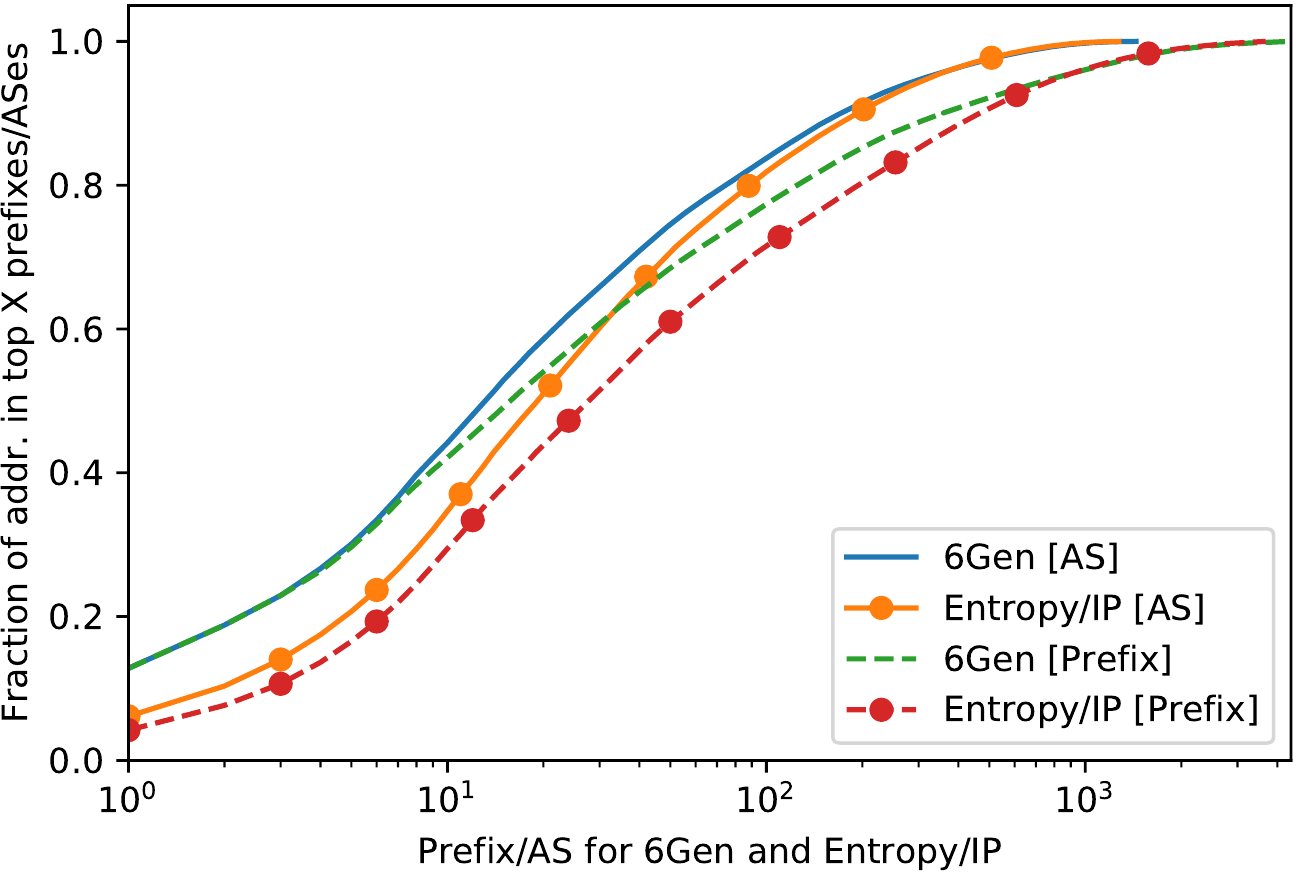}
    \caption{Prefix and AS distribution for responsive addresses generated with \sixgen and \eip.}
    \label{fig:pfxas6geneip}
\end{figure}

Finally, we compare ASes and prefixes of responsive addresses for both tools.
\sixgen discovers responsive hosts in \num{1442} ASes, while \eip does in \num{1275} ASes.
Interestingly, responsive hosts in 384 ASes are found by only one of the tools, \ie either \sixgen or \eip.
In \Cref{fig:pfxas6geneip}, we show the prefix and AS distributions of responsive hosts.
\eip's distribution is a bit less top-heavy compared to \sixgen's, where the top 2 responsive ASes make up almost \sperc{20} of all addresses.
Although there is some overlap in the top 5 ASes, \sixgen features more ISPs, like Sky Broadband, Google Fiber, and Xs4all Internet.
In contrast, \eip's top ASes contain more CDNs and Internet services.

To summarize, \sixgen and \eip find few overlapping responsive addresses, but mostly in overlapping ASes.
The services offered by these hosts differ considerably.
Therefore, both tools have their advantages in finding specific addresses and populations.
We suggest running both tools to maximize the number of found responsive addresses.
\section{rDNS as a Data Source}\label{sec:rdns}

In addition to the sources described in \Cref{sec:sources}, we investigate the usefulness of IPv6 rDNS entries for active measurements.
As shown by previous work, rDNS walking can be a source for IPv6 addresses~\cite{fiebig2017something,fiebig2018rdns}.
While IPv6 rDNS addresses were used to, \eg find misconfigured IPv6 networks~\cite{borgolte2018enumerating}, we are not aware of studies evaluating overall responsiveness. 
Since walking the rDNS tree to harvest IPv6 addresses is a large effort and puts strain on important Internet infrastructure, we classify this source as   ``semi-public'', compared with sources such as the Alexa Top \sm{1} list, which is available for download. 

We use IPv6 rDNS data provided by Fiebig \etal \cite{fiebig2018rdns} to perform active measurements and compare the results against other hitlist sources. Analyzing the overlap and structure of IPv6 addresses obtained from rDNS, we find a very small intersection with our \hitlist.
Of the \SI{11.7}{\mn} addresses from rDNS, \SI{11.1}{\mn} are new.
The prefix distribution of rDNS and hitlist addresses is quite similar, as shown in \Cref{fig:pfxasrdns}.
The AS distribution is even more balanced for rDNS addresses compared to the hitlist.
Therefore, the addition of rDNS data to the hitlist input would not introduce a bias at the prefix or AS level. %

Next, we perform active measurements to compare the response rate of the rDNS population to the hitlist population.
Before the active measurement, we filter \SI{2.1}{\mn} unrouted addresses and \SI{13.1}{\kn} addresses residing in aliased prefixes (see \Cref{sec:probing:sub:aliasprefixdetection}) from the rDNS addresses.
The response rate for the hitlist with only non-aliased prefixes is generally similar to the response rate of rDNS.
The rDNS ICMP response rate is higher: \SI{10}{\percent} compared to the hitlist's \SI{6}{\percent}.
On the other hand, we receive slightly fewer HTTP(S) responses for rDNS, at \SI{2}{\percent} (\SI{1}{\percent}) against the hitlist's \SI{3}{\percent} (\SI{2}{\percent}).

To ensure that responding rDNS addresses are not mostly client addresses, we first analyze the top ASes. As can be seen in \Cref{tab:rdnsas}, the top responsive ASes in the rDNS data are hosting and service providers, \ie mostly servers (especially in the TCP/80 measurement).
\begin{figure}[bt]%
	\includegraphics[width=\columnwidth]{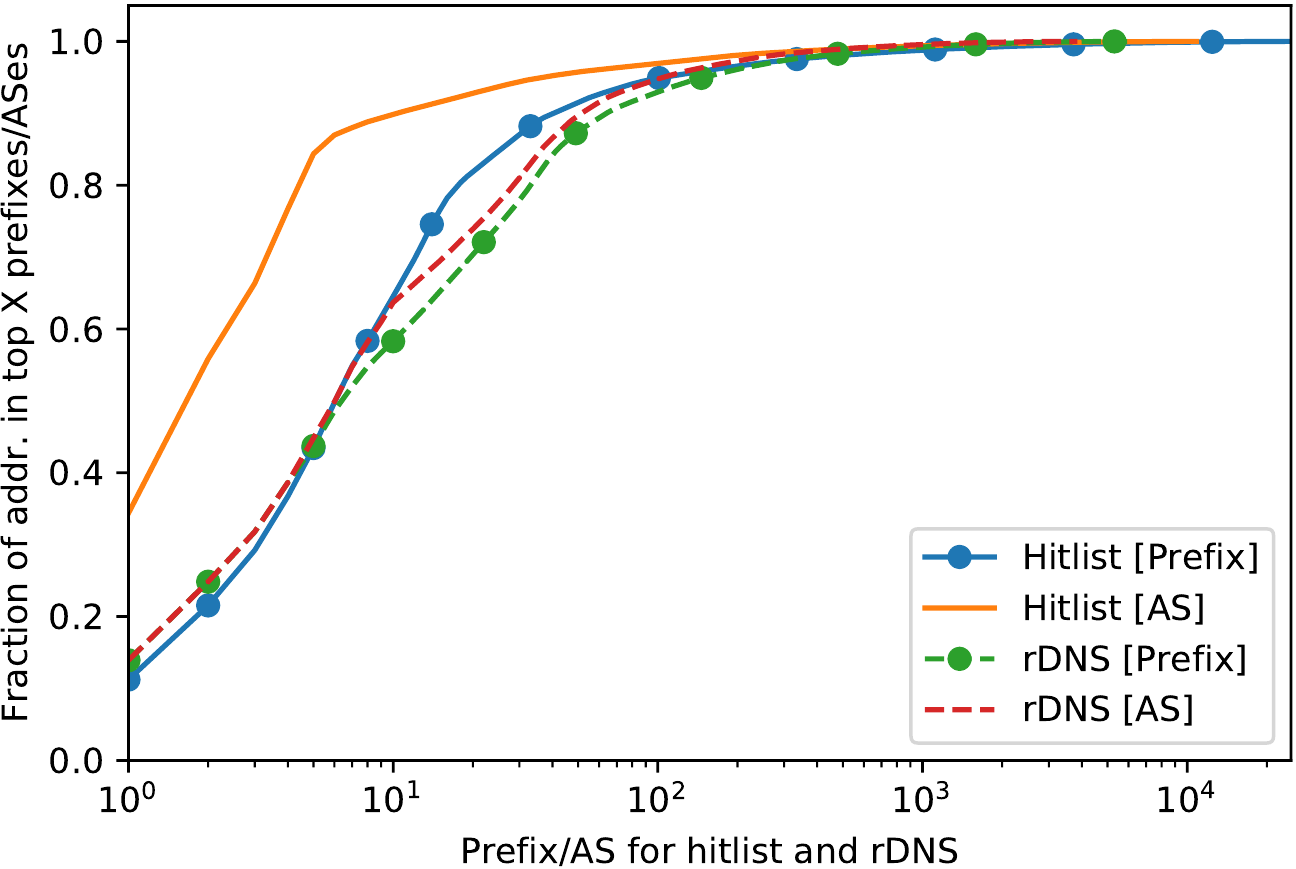}
	\caption{Prefix and AS distribution for hitlist and rDNS input data.}
	\label{fig:pfxasrdns}
\end{figure}
\begin{table}[bt]%
	\centering
	\caption{Top 5 rDNS ASes in input and responsive via ICMP and TCP/80.}
	\label{tab:rdnsas}
	\resizebox{\columnwidth}{!}{
		\begin{tabular}{ll@{\hskip .5\tabcolsep}rl@{\hskip .5\tabcolsep}rl@{\hskip .5\tabcolsep}r}
			\toprule
			\# & \multicolumn{2}{c}{Input} & \multicolumn{2}{c}{ICMP} & \multicolumn{2}{c}{TCP/80} \\
			\midrule
			1 & Comcast & \sperc{12.5} & Online S.A.S. & \sperc{19.6} & Google & \sperc{12.8} \\
			2 & AWeber & \sperc{10.2} & Sunokman & \sperc{17.8} & Hetzner & \sperc{10.1} \\
			3 & Yandex & \sperc{9.8} & Latnet Serviss & \sperc{8.7} & Freebit & \sperc{6.8} \\
			4 & Belpak & \sperc{6.2} & Yandex & \sperc{7.9} & Sakura & \sperc{6.5} \\
			5 & Sunokman & \sperc{6.1} & Salesforce & \sperc{5.3} & TransIP & \sperc{5.0} \\
			\bottomrule
		\end{tabular}
	}
\end{table}
Next, we look for IPv6 SLAAC's distinct \texttt{ff:fe} sequence and evaluate the hamming weight of IIDs for responsive rDNS addresses, as an additional indicator for clients.
We find between \SI{6}{\percent} and \SI{9}{\percent} SLAAC addresses with \texttt{ff:fe}.
The IID hamming weight (\ie number of bits set to 1) can be used to infer the presence of clients with privacy extensions enabled \cite{Gasser2016a}.
The rDNS IID hamming weight does not suggest that the rDNS set contains a large client population, especially for TCP/80, where \SI{60}{\percent} of addresses have a hamming weight of six or smaller.

To conclude, the responsive part of the rDNS data source adds a balanced set of IPv6 addresses.
We therefore suggest adding rDNS data as input to the IPv6 hitlist.
\section{Client IPv6 addresses}\label{sec:client}

The majority of IPv6 addresses that we have collected so far belong to servers or routers rather than clients. 
In this section, we use crowdsourcing platforms to collect client IP addresses.
Our aim is to begin studying the following questions: Does crowdsourcing serve as an adequate source of residential IPv6 addresses, and might we be able to use addresses gathered through crowdsourcing in IPv6 hitlists?
We refer the reader to~Section \ref{sec:ethical} describing the ethical consideration of this study.

\subsection{Experimental Setup} 

To investigate our approach of collecting IPv6 client addresses, we use two crowdsourcing platforms and leverage the \textit{test-ipv6.com}~\cite{test} code to gather IPv6 addresses.

We set up a web server and integrate it into the crowdsourcing environment to operate similarly to the study by Huz \etal \cite{huz2015experience}. %
We use Amazon Turk (Mturk) \cite{mturk} and Prolific Academic (ProA) \cite{proa} to run our experiments 
and allocate a budget of \sdollar{150} per platform.
We add a limitation for only one submission per user per platform and select the minimum amount of money that can be set for each assignment: \sdollar{0.01} for Mturk and \sdollar{0.12} for ProA. 
We run the experiment between April 23 and May 23, 2018. 
During this time, \num{5781} users participate from Mturk, compared to \num{1186} from ProA. We make our setup and source code available to the community~\cite{cs:ipv6}. 

\subsection{Crowdsourcing Participants}
 
\Cref{cs:dist:m} shows the distribution of measurements per platform, including AS \cite{pyasn} and country mappings \cite{maxmind}.
We find about \sperc{31} of Mturk and \sperc{20.6} of ProA users with IPv6 enabled.
One reason for Mturk's higher IPv6 ratio might be their customer base:
Mturk is more popular in the US and India~\cite{spoof,huz2015experience}, both countries with considerable IPv6 adoption~\cite{adoption}.

Moreover, \sperc{31.5} of IPv6 ASes are overlapping between platforms, although  we do not find any common addresses.

A large part of our IPv6 clients participate from a small number of ISPs.
The top 3 ASes are Comcast (\sperc{31.1}) and AT\&T (\sperc{13.2}),
and the Indian ISP Reliance (\sperc{7.8}).
Comparing IPv6 clients with IPv4 clients we find the latter to be more diverse, where the top 5 providers constitute only \sperc{30} of all IPv4 clients in our dataset.  

\begin{table}[ht]
\centering
\caption{Client distribution in crowdsourcing study.}
\label{cs:dist:m}
\begin{tabular}{lrrrrrr}
  \toprule
        & IPv4 & IPv6 & ASes$_4$ & ASes$_6$ & \#cc$_4$ & \#cc$_6$ \\
    \midrule
    Mturk & \num{5707} & \num{1787} & \num{842}  & \num{73}   & \num{93}  & \num{22}  \\
    ProA  & \num{1176} & \num{245}  & \num{272}  & \num{48}   & \num{33}  & \num{21} \\
     \midrule
     Unique  &  \num{6862} & \num{2032}  & \num{983}  & \num{92}   & \num{98}  & \num{29} \\
  \bottomrule
\end{tabular}
\end{table}

\subsection{Client Responsiveness}
Once the user submits the results, we send an ICMPv6 echo request %
and traceroute to each IPv6 address every 5 minutes. %

We find that only \num{352} (\sperc{17.3}) of IPv6 addresses respond to at least one ICMPv6 echo request.
The majority of IPv6 addresses we gathered from residential networks do not respond to these probes. %

To investigate whether the low response rate was an artifact of the devices (\eg privacy extension address cycling, users disconnecting), or network policy, we locate the set of RIPE Atlas probes situated in the same ASes as our crowdsourced client addresses.
RIPE Atlas probes %
generally respond to echo requests they receive. %

We select \num{1398} RIPE Atlas probes with IPv6 connectivity inside these ASes and, having confirmed they were online, send traceroutes to those probes.
As many as \num{641} (\sperc{45.8}) probes respond to our queries. 
Since RIPE Atlas probes will respond by design, these \sperc{45.8} indicate an upper bound of possible crowdsourcing responses.
It is likely that users' systems are running local firewalls which will reduce the response rate further. 
This is indicative only, and deserves further study.

We also find that for \sperc{20} of clients, the last responsive hop is different from the destination AS, which indicates filtering by ISPs. 

The low responsiveness from the RIPE Atlas probes and crowdsourcing clients suggests inbound filtering. RFC 7084~\cite{rfc7084} leaves it open for the user to decide to either have an ``outbound only'' or an ``open'' configuration, allowing all internally and externally initiated ICMPv6 connections for IPv6 CPEs. Future work is needed to understand the reasons for the deployment of ``outbound only'' policy for ICMPv6 in residential networks.

Client addresses that respond to our ICMPv6 requests are likely to be less stable than server IP addresses. 
Only \num{7} IPv6 addresses from the total of \num{352} responsive addresses remain active for the entire month.
\sperc{19} of IPv6 addresses are active for less than an hour, while \sperc{39.4} of addresses are active for \num{8} hours or less. %
Moreover, addresses with dynamic behavior \ie appearing and disappearing multiple times during the study, had a mean uptime of  
approximately \num{8} hours and median uptime of \num{3} hours per day.

We conclude that crowdsourced addresses provide additional targets for IPv6 client studies.
Only a fraction of them are, however, responsive to incoming probes. Future work is needed to get more representative data and to understand the extent to which filtering is performed. 
Finally, active measurements targeting crowdsourcing addresses need to be performed swiftly after address collection, as the responsive client population is quickly shrinking.

\section{Measurement practices} \label{sec:practices}

In our study we follow measurement best practices by conducting scans in an ethical way and publishing data and code for reproducibility in research.

\subsection{Ethical Considerations}\label{sec:ethical}

Before conducting active measurements we follow an internal multi-party approval process, which incorporates proposals by Partridge and Allman \cite{partridge2016ethical} and Dittrich \etal \cite{dittrich2012menlo}.
We assess whether our measurements can induce harm on individuals in different stakeholder groups.
As we limit our query rate and use conforming packets, it is unlikely that our measurements will cause problems on scanned systems.
We follow scanning best practices \cite{Durumeric2013} by maintaining a blacklist and using dedicated servers with informing rDNS names, websites, and abuse contacts.
During our six months of active scans, we received four emails asking for more information on the conducted measurements.

Prior to deploying the crowdsourcing study we discussed the subject with the university ethics committee.
We agreed with the committee that IPv6 addresses collected in the crowdsourcing study will be excluded from public datasets.
In addition, we informed participants that they could opt out of the active probing.
The ethics committee concluded that our experiments did not constitute research on human subjects, as we have no reasonable way of mapping collected IPv6 addresses to individuals, and gave the study formal permission. %

\subsection{Reproducible Research}

To encourage reproducibility in network measurement research \cite{AcmArtifacts,reproduc2017}, we make data, source code, and analysis tools publicly available \cite{tubib}.
This includes \zesplot \cite{zesplot}, entropy clustering \cite{entropyclustering}, new \eip generator \cite{eip-generator}, crowdsourcing documentation \cite{cs:ipv6}, and results from daily runs of the IPv6 hitlist, including the list of aliased prefixes \cite{ipv6hitlist}.
Some of the figures in this paper are clickable, offering additional insights, in-depth analyses, and interactive graphs.
This data can serve as a valuable starting point for future IPv6 studies.
\section{Discussion}
\label{sec:discussion}

In this section we summarize challenges in generating hitlists and discuss lessons learned for future work.

\textbf{Time-to-Measurement:} We collect IPv6 addresses from a variety of sources containing server, router, and client addresses.
Our analysis shows that server IPv6 addresses are more responsive and stable in comparison to CPE and client devices.
As a result, when using an IPv6 hitlist as an input for a specific measurement study, researchers need to consider the time-to-measurement: client devices need to be measured within minutes to obtain sensible response rates, whereas servers remain responsive over weeks.

\textbf{Hitlist Tailoring:} To reduce bias when conducting IPv6 measurements, we generally advise to strive for an evenly balanced hitlist across prefixes and ASes, and to remove addresses in aliased prefixes.
Depending on the research goal, researchers can pivot from an even address distribution to a stronger focus on certain address types (\eg HTTPS web servers).
Depending on the type of study it may also be desirable to include or exclude specific data sources.
For example, studies analyzing hosting providers can use server addresses, while for residential networks, researchers can focus on sources containing mainly CPE and client addresses.

\textbf{Unresponsive Addresses:} While our focus on responsive addresses is reasonable for a hitlist which is used as direct input for a measurement study, there might be scenarios where also unresponsive addresses could be of value.
Unresponsive addresses can be used to understand addressing schemes inside a prefix.
They can also be used as an input for address learning algorithms (\eg \eip or \sixgen) which might then output responsive addresses.

\textbf{IPv6 Hitlist Service:} In order to help future IPv6 measurements we provide daily IPv6 hitlists and a list of aliased prefixes at:\\
\centerline{\url{https://ipv6hitlist.github.io}}\\
Providing an easy-to-use hitlist service has several advantages for IPv6 research:
\one While conducting measurements on all known IPv6 addresses is possible in one-off measurements, high-frequency periodic measurements might require a focus on a subset of responsive addresses.
We provide lists of responsive addresses which can be directly used in IPv6 measurement studies.
\two With the increasing adoption of IPv6, the number of all publicly known IPv6 addresses is bound to increase as well, making it increasingly challenging to perform a full sweep of all known addresses in the future.
\three Reducing the number of measurement targets for the multitude of IPv6 measurement studies is considered good Internet citizenship~\cite{klwr-tbicr-16b}.
\four By providing historical data we enable other researchers to study the evolution of IPv6 deployment.
\section{Conclusion}

In this work, we leveraged a multitude of sources to create the largest IPv6 \hitlist to date, containing more than \sm{50} addresses.
We found that about half of these addresses reside in aliased prefixes and showed, using clustering, that there are only six prevalent IPv6 addressing schemes.
Using longitudinal measurements we identified protocols and sources which are less stable over time.
We used and extended state-of-the-art tools to generate new addresses, finding that they provide complementary address sets.
We will keep running daily IPv6 measurements to provide valuable hitlists. %

\noindent\textbf{Acknowledgments:} We would like to thank Austin Murdock for providing a pre-release version of \sixgen and Tobias Fiebig for the IPv6 rDNS data.
The authors would like to thank the anonymous reviewers and our shepherd Robert Beverly for their valuable feedback.
This work was partially funded by the German Federal Ministry of Education and Research under the projects X-Check\\ (16KIS0530), DecADe (16KIS0538), and AutoMon (16KIS0411).

\mathchardef\UrlBreakPenalty=100

\bibliographystyle{plainurl}
\bibliography{clustersintheexpanse}

\end{document}